\documentclass[aoas,MSNbibl,nameyear,seceqn,dvips]{arximspdf}
\usepackage[linesnumbered,lined,ruled]{algorithm2e}
\usepackage{graphicx}
\doi{10.1214/11-AOAS528}
\volume{6}
\issue{2}
\pubyear{2012}
\firstpage{561}
\lastpage{600}

\makeatletter

\newcommand{\eqref}[1]{(\ref{#1})}

\newcommand{\1}{\mathbf{1}}
\newcommand{\E}{\mathcal{E}}
\newcommand{\G}{\mathcal{G}}
\newcommand{\HZ}{\mathbf{H}_0}
\newcommand{\HO}{\mathbf{H}_1}
\newcommand{\LL}{\mathcal{L}}
\newcommand{\N}{\mathcal{N}}
\newcommand{\RR}{\mathbb{R}}
\newcommand{\V}{\mathcal{V}}
\newcommand{\cF}{F_0} 
\newcommand{\ncF}{F} 
\def\argmin{\arg\min}
\newcommand{\Diag}{\operatorname{Diag}}
\newcommand{\Indic}{\mathrm{I}}
\makeatother

\begin{document}
\begin{frontmatter}

\title{More power via graph-structured tests for differential
expression of gene networks}
\runtitle{Graph-structured tests for differential expression}

\begin{aug}
\author[A]{\fnms{Laurent} \snm{Jacob}\ead[label=e1]{laurent@stat.berkeley.edu}},
\author[B]{\fnms{Pierre} \snm{Neuvial}\ead[label=e2]{pierre.neuvial@genopole.cnrs.fr}}
\and
\author[C]{\fnms{Sandrine} \snm{Dudoit}\corref{}\ead[label=e3]{sandrine@stat.berkeley.edu}}
\runauthor{L. Jacob, P. Neuvial and S. Dudoit}
\affiliation{University of California, Berkeley, University of California, Berkeley, and~Universit\'e d'\'Evry Val d'Essonne}
\address[A]{L. Jacob\\
Department of Statistics\\
University of California, Berkeley\\
Berkeley, California 94720-7360\\
USA\\
\printead{e1}} 
\address[B]{P. Neuvial\\
Department of Statistics\\
University of California, Berkeley\\
Berkeley, California 94720-7360\\
USA\\
and\\
Laboratoire Statistique et G\'enome\\
Universit\'e d'\'Evry Val d'Essonne -- \\
\quad UMR CNRS 8071 -- USC INRA\\
France\\
\printead{e2}}
\address[C]{S. Dudoit\\
Department of Statistics\\
Division of Biostatistics\\
University of California, Berkeley\\
Berkeley, California 94720-7360\\
USA\\
\printead{e3}}
\end{aug}

\received{\smonth{11} \syear{2010}}
\revised{\smonth{10} \syear{2011}}

%
\begin{abstract}
We consider multivariate two-sample tests of means, where the
location shift between the two populations is expected to be related
to a known graph structure. An important application of such tests
is the detection of differentially expressed genes between two
patient populations, as shifts in expression levels are expected to
be coherent with the structure of graphs reflecting gene properties
such as biological process, molecular function, regulation or
metabolism. For a fixed graph of interest, we demonstrate that
accounting for graph structure can yield more powerful tests under
the assumption of smooth distribution shift on the graph. We also
investigate the identification of nonhomogeneous subgraphs of a
given large graph, which poses both computational and multiple
hypothesis testing problems. The relevance and benefits of the
proposed approach are illustrated on synthetic data and on breast
and bladder cancer gene expression data analyzed in the context of
KEGG and NCI pathways.
\end{abstract}

%

\end{frontmatter}

\section{Introduction}
\label{sec:intro}

The detection of differentially expressed (DE) genes, that is, genes
whose expression levels change between two (or more) experimental
conditions, remains a major challenge in biology and medicine,
especially in the context of cancer studies. For example, the
identification of DE genes between breast cancer patients that are
sensitive or resistant to tamoxifen can help understand resistance
mechanisms to this drug and eventually improve breast tumor
treatment [\citet{Loi2008Predicting}]. Similarly, finding DE genes
between low-grade, noninvasive or more aggressive bladder tumors may
help understand the disease better and ultimately improve its
diagnosis and treatment [\citet{Stransky2006Regional}]. The application
of the methods developed in this paper will be illustrated on the
data sets from the above two papers.

However, the detection of a change in gene expression levels among a~large gene list is a difficult problem from a statistical perspective,
and lists of differentially expressed genes are generally hard to
interpret, as they focus on the level of genes instead of the level of
molecular functions. In such a context, expression data from
high-throughput microarray and sequencing assays gain much in
relevance from their association with graph-structured prior
information on the genes, {for example}, Gene Ontology (GO;
\url{http://www.geneontology.org}), Kyoto Encyclopedia of Genes and
Genomes (KEGG; \url{http://www.genome.jp/kegg}) or NCI Pathway
Integration Database (NCI graphs; \url{http://pid.nci.nih.gov}). Most
approaches to the joint analysis of gene expression data and gene
graph data involve two distinct steps. First, tests of differential
expression are performed separately for each gene. Then, these
univariate (gene-level) testing results are extended to the level of
gene sets, {for example}, by assessing the over-representation
of DE
genes in each set based on $p$-values for Fisher's exact
test\footnote{Sometimes referred to as a hypergeometric test in the
bioinformatics literature.} (or a $\chi^2$ approximation thereof)
adjusted for multiple testing [\citet{Beissbarth2004GOstat}] or based on
permutation adjusted $p$-values for weighted Kolmogorov--Smirnov-like
statistics [\citet{Subramanian2005Gene}]. Another family of methods
directly performs multivariate tests of differential expression for
groups of genes, {for example}, Hotelling's
$T^2$-test [\citet{Lu2005Hotelling}]. It is
known [\citet{Goeman2007Analyzing}] that the former family of approaches
can lead to incorrect interpretations, as the sampling units for the
tests in the second step become the genes (as opposed to the patients)
and these are expected to have strongly correlated expression
measures. This fact suggests that direct multivariate testing of gene
set differential expression is more appropriate than posterior
aggregation of individual gene-level tests. On the other hand, while
Hotelling's $T^2$-statistic is known to perform well in small
dimensions, it loses power very quickly with increasing
dimension [\citet{Bai1996Effect}], essentially because it is based on
the inverse of the empirical covariance matrix which becomes
ill-conditioned. Additionally, such direct multivariate tests on
unstructured gene sets do not take advantage of information on gene
regulation or other relevant biological properties. An increasing
number of regulation networks are becoming available, specifying, for
example, which genes activate or inhibit the expression of which other
genes. If it is known that a particular gene in a tested gene set
activates the expression of another, then one expects the two genes to
have coherent (differential) expression patterns, {for
example}, higher
expression of the first gene in resistant patients should be
accompanied by higher expression of the second gene in these patients.
Accordingly, the first main contribution of this paper is to propose
and validate multivariate test statistics for identifying differential
expression patterns (or, more generally, shifts in distribution) that
are coherent with a given graph structure.\vadjust{\goodbreak}

Next, given a large graph and observations from two data generating
distributions on the graph, a more general problem is the
identification of smaller nonhomogeneous subgraphs, {that is},
subgraphs on which the two distributions (restricted to these
subgraphs) are significantly different. This is very relevant in the
context of tests for gene set differential expression: given a large
set of genes, together with their known regulation network, or the
concatenation of several such overlapping sets, it is important to
discover novel gene sets whose expression changes significantly between
two conditions. Currently-available gene sets have often been defined
in terms of other phenomena than that under study and physicians may
be interested in discovering sets of genes affecting in a concerted
manner a specific phenotype. Our second main contribution is
therefore to develop algorithms that allow the exhaustive testing of
all the subgraphs of a large graph, while avoiding one-by-one
enumeration and testing of these subgraphs and accounting for the
multiplicity issue arising from the vast number of subgraphs.

As the problem of identifying variables or groups of variables which
differ in distribution between two populations is closely related to
supervised learning, our proposed approach is similar to several
learning methods. \citet{Rapaport2007Classification} use filtering in
the Fourier space of a graph to train linear classifiers of gene
expression profiles whose weights are smooth on a~gene
network. However, their classifier enforces global smoothness on the
large regularization network of all the genes, whereas we are
concerned with the selection of gene sets with locally-smooth
expression shift between populations. In \citet{Jacob2009Group} and
\citet{Obozinski2011Group}, sparse learning methods are used to build a
classifier based on a small number of gene sets. While this approach
leads in practice to the selection of groups of variables whose
distributions differ between the two classes, the objective is to
achieve the best classification performance with the smallest possible
number of groups. As a result, correlated groups of variables are
typically not selected. Other related work
includes \citet{Fan1998Test}, who proposed an adaptive Neyman test in
the Fourier space for time series. However, as illustrated below in
Section~\ref{sec:synexp}, direct translation of the adaptive Neyman
statistic to the graph case is problematic, as assumptions on Fourier
coefficients which are true for time series do not hold for graphs.
In addition, the Neyman statistic converges very slowly toward its
asymptotic distribution and the required calibration by bootstrapping
renders its application to our subgraph discovery context
difficult. By contrast, other methods do not account for shift
smoothness and try to address the loss of power caused by the poor
conditioning of the $T^2$-statistic by applying it after
dimensionality reduction [\citet{Ma2009Identification}] or by omitting
the inverse covariance matrix and adjusting instead by its
trace [\citet{Bai1996Effect}, \citet{Chen2010A}] or using a diagonal
estimator of
the covariance
matrix [\citet{Srivastava2008Test}, \citet{Srivastava2009Test}].
\citet{Lopes2011More}
recently proposed a testing procedure based on random projection of
the data in a lower dimension space, and showed that it was
asymptotically more powerful
than \citet{Bai1996Effect}, \citet{Chen2010A} and
\citet{Srivastava2008Test} in the presence of correlation and when the
spectrum of the covariance matrix decays fast
enough. \citet{Vaske2010Inference} recently proposed DE tests, where a
probabilistic graphical model is built from a gene network. However,
this model is used for gene-level DE tests, which then have to be
combined to test at the level of gene sets. Several approaches for
subgraph discovery, like that of \citet{Ideker2002Discovering}, are
based on a heuristic to identify the most differentially expressed
subgraphs and do not amount to testing exactly all possible
subgraphs. Concerning the discovery of distribution-shifted subgraphs,
\citet{Vandin2010Algorithms} propose a graph Laplacian-based testing
procedure to identify groups of interacting proteins whose genes
contain a large number of mutations. Their approach does not enforce
any smoothness on the detected patterns (smoothness is not necessarily
expected in this context) and the graph Laplacian is only used to
ensure that very connected genes do not lead to spurious
detection. The Gene Expression Network Analysis (GXNA) method of
\citet{Nacuetal07} detects differentially expressed subgraphs based on
a greedy search algorithm and gene set DE scoring functions that do
not account for the graph structure.

The rest of this paper is organized as
follows. Section~\ref{sec:smooth} explains how to build a
lower-dimension basis in which to apply the multivariate test of
means. Section~\ref{sec:test} presents our graph-structured two-sample
test statistic and states results on power gain for smooth-shift
alternatives. Section~\ref{sec:discovery} describes procedures for
systematically testing (without fully enumerating) all possible
subgraphs of a large graph. Section~\ref{sec:synexp} presents results
for synthetic data and Section~\ref{sec:geneexp} on breast and bladder
cancer gene expression data sets analyzed in the light of pathways from
the KEGG and NCI databases. Section~\ref{sec:soft} presents softwares
implementing the proposed methods. Finally,
Section~\ref{sec:discussion} summarizes our findings and outlines
ongoing work.

Although this work is motivated by the specific question of
differential expression testing of gene networks, our proposed
structured two-sample test of means on a graph and our nonhomogeneous
subgraph discovery algorithm can actually be used in any situation
where one searches for differences between two populations that are
expected to be coherent with a known graph structure. Therefore, our
methodological contributions in Sections~\ref{sec:test}
and~\ref{sec:discovery} are presented in the general context of
two-sample tests on graphs.

\section{Graph-based dimensionality reduction}
\label{sec:smooth}

\begin{figure}

\includegraphics{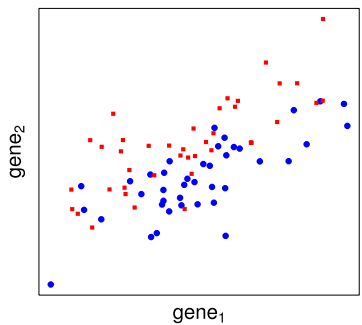}

\caption{Synthetic example of the joint distribution of the
expression measures of two genes in two patient populations. The
color and shape of the plotting symbols indicate the patient group
and the $x$- and $y$-axes correspond to the expression measures of the
first and second gene, respectively.}
\label{fig:mtwosample}
\end{figure}

As stated in the \hyperref[sec:intro]{Introduction}, each of the two main paradigms for
testing differential expression of a gene set have their
limitations. Two-step methods generally do not directly test
the\vadjust{\goodbreak}
existence of a mean shift between two multivariate distributions
[\citet{Goeman2007Analyzing}]. The second step, which often treats the
\textit{genes} as the sampling units, renders the interpretation of
$p$-values problematic and may lead to a large loss of power or Type I
error control when sets of genes have correlated
expression. Multivariate statistics, on the other hand, allow a direct
formulation of and solution to the testing question: the sampling
units are vectors of gene expression measures ({e.g.},
corresponding to patients) and the question is whether two such sets
of random vectors are likely to have arisen from distributions with
equal means. Figure~\ref{fig:mtwosample} illustrates another classical
advantage of multivariate approaches: genes taken individually may
have extremely small mean shifts between two populations, although
their joint distributions clearly differ between the two
populations. Here, again, this phenomenon typically happens for sets
of genes whose expression measures are correlated, which is not
unlikely for pathways or annotated gene sets.

Unfortunately, with moderate sample sizes, multivariate statistics
lose power quickly in a high dimension. If some type of side information
is available regarding particular properties of the expression shift,
a possible approach to get the best of both worlds would be to: (1)
project the vectors of covariates in a new space of
\textit{lower dimension} that preserves the distribution shift,
{that is}, the distance between the expression measures of the two
groups, and (2) apply the multivariate statistic in this new
space. One could thus perform the appropriate multivariate test, while
avoiding the loss of power caused by the high-dimensionality of the
original covariate space.

A possible source of information about the expression shift is the
growing number of available gene networks. Indeed, while the
difference in mean expression between two groups of patients may not
be entirely coherent with an existing network ({e.g.}, because of
noise in the data, errors in the annotation, or inappropriateness of
the chosen network for the biological question of interest), it is
reasonable to expect that this shift will not be entirely
contradictory with the given graph structure. For example, repressed
genes should be more connected to other repressed genes than to
overexpressed genes. Given this assumption, we intend to build a space of
lower dimension than the original gene space, but which preserves most
of the distribution shift between the two populations.

More precisely, consider a network of $p$ genes, represented by a
graph $\G= (\V,\E)$, with $|\V| = p$ nodes and edge set
$\E$. Let $\delta\in\RR^p$ denote the mean shift, {that
is}, the vector
of differences in mean expression measures for these~$p$ genes between
the two populations of interest. Suppose we expect the shift~$\delta$
to be
coherent with the graph $\G$, in the sense that it has low energy
$E_\G(\delta)$ for a~particular energy function $E_\G$ defined on
$\G$. Then, we wish to build a~space of lower dimension $k \ll p$
capturing most of the low energy functions. To this end, we start by
finding the function that has the lowest possible energy, then the
function that has lowest possible energy in the orthogonal space of
the first one, up to the $k$th function with lowest energy in the
orthogonal subspace of the first $k-1$ functions. That is, for each
$i\leq k$, we define
%
\begin{equation}
\label{eq:smoothBasis}
u_i=
\cases{
\displaystyle\mathop{\argmin}_{f\in\RR^p}E_\G(f)\vspace*{2pt}\cr
\mbox{such that } u_i \perp u_j, j < i.}
\end{equation}

If $E_\G$ is a positive semi-definite quadratic form $E_\G(\delta) =
\delta^\top Q_\G\delta$, for some positive semi-definite matrix
$Q_\G
= U\Lambda U^\top$, where $U$ is an orthogonal matrix and $\Lambda$ a
diagonal matrix with elements $\lambda_i, i=1,\ldots,p$, then the
solution to equation~\eqref{eq:smoothBasis} is given by the $k$
eigenvectors of $Q_\G$ corresponding to the smallest $k$ eigenvalues. It
is easy to check that these eigenvalues are the energies of the
corresponding functions $u_i$, {that is}, $E_\G(u_i) = \lambda_i$.

Different choices of $Q_\G$ lead to different notions of coherence of
the expression shift with the network. A classical choice is the
\textit{graph Laplacian} $\LL$. Suppose $\G$ is an \textit{undirected}
graph with adjacency matrix $A$, with $a_{ij}=1$ if and only if
$(i,j)\in\E$ and $a_{ij}=0$ otherwise, and degree matrix $D =
\Diag(A \1)$, where $\1$ is a unit column-vector,
$\Diag(x)$ is the diagonal matrix with diagonal $x$ for any vector
$x$, and $D_{ii} = d_i$. The Laplacian matrix\vspace*{1pt} of $\G$ is then
typically defined as $\LL= D - A$ or $\LL_{\mathrm{norm}} = I -
D^{-{1}/{2}}AD^{-{1}/{2}}$ for the normalized version, leading
to energies $\sum_{i,j\in\V}(\delta_i - \delta_j)^2$ and
$\sum_{i,j\in\V}(\frac{\delta_i}{\sqrt{d_i}} -
\frac{\delta_j}{\sqrt{d_j}})^2$, respectively. Note that, in
this case, the Laplacian matrix $\LL$, energy $E$ and basis functions
$u_i$ extend the classical Fourier analysis of functions on Euclidean
spaces to functions on graphs, by transferring the notions of Laplace
operator, Dirichlet energy and Fourier basis,
respectively~[\citet{Evans1998Partial}].

More generally, any positive semi-definite matrix can be chosen. In
the case of gene regulation networks, we do not necessarily expect as
strong a~coherence as that corresponding to the Dirichlet energy
defined by the graph Laplacian, since some of the annotated
interactions may not be relevant in the studied context and some
antagonist interactions may cancel each other. For example, if a gene
is activated by two others, one who is underexpressed and the other
overexpressed, we may observe no change in the expression of the
gene, but a nonzero Dirichlet energy $\sum_{i,j\in\V}(\delta_i -
\delta_j)^2$. Additionally, for applications like structured
gene set differential expression detection, one may use negative
weights for edges that reflect a negative correlation between two
variables, {for example}, a gene $i$ whose expression inhibits the
expression of another gene $j$. In this case, a small variation of the
shift on the edge between $i$ and $j$ should correspond to a small
$|\delta_i + \delta_j|$. This can be achieved in the same formalism by
simply considering a \textit{signed} version of the adjacency matrix $A$,
{that is}, $a_{ij} = 1$ if gene $i$ activates gene $j$ and $-1$ if
it inhibits gene $j$. A signed version of the graph Laplacian is then
$\LL_{\mathrm{sign}} = D - A$, where $D = \Diag(|A| \1)$ is
the degree matrix and $|A|$ denotes the entry-wise absolute value of
$A$. Note that such a signed Laplacian was used as a penalty for
semi-supervised learning in \citet{Goldberg2007Dissimilarity}.

In the context of this work, we, moreover, consider \textit{directed}
graphs $\G= (\V,\E)$, where the edge set $\E$ consists of
ordered pairs of nodes. The adjacency matrix $A$ may be asymmetric,
with entries $a_{ij} \neq0$ if and only if $(i,j)\in\E$,
{that is},
there is an (directed) edge pointing from node $v_i$ to node $v_j$. We
then use the following energy function:
%
%
\begin{equation}
\label{eq:energy}
E_\G(\delta) = \sum_{i:d_i^{-}\neq0}^p\biggl(\delta_i -
\frac{1}{d_i^{-}}\sum_{(j,i)\in\E}a_{ji}\delta_j\biggr)^2,
\end{equation}
where $d_i^{-}   \stackrel{\Delta}{=} \sum_{j=1}^p|a_{ji}|$ is the
indegree of node $v_i$, {that is}, the number of directed edges
pointing from any node to $v_i$. According to this definition, an expression
shift will have low energy if the difference in mean expression of any
given gene between the two populations is similar to the (signed)
average of the differences in mean expression for the genes that
either activate or inhibit it.

It is immediate to check that $E_\G(\delta)\!=\!\delta^\top M_\G
\delta$, with $M_\G\!\stackrel{\Delta}{=}\!(\tilde{I}\!-\!D_{-}^{-1}A^\top)^\top(\tilde{I}\!- D_{-}^{-1}A^\top)$, where $D_{-}
\stackrel{\Delta}{=} \Diag((d^{-}_i)_{i=1,\ldots,p})$ is
the matrix of indegrees, $\tilde{I} \stackrel{\Delta}{=}
\Diag((\Indic(d^{-}_i \neq0))_{i=1,\ldots,p})$ is a
modification of the identity matrix where diagonal elements
corresponding to nodes with zero indegree are set to zero, and the
value of the indicator function $\Indic$ is $1$ if its argument is
true and zero otherwise. Note that a very similar function was used in
the context of
regularized supervised learning by \citet{Sandler2008Regularized}.

\begin{figure}

\includegraphics{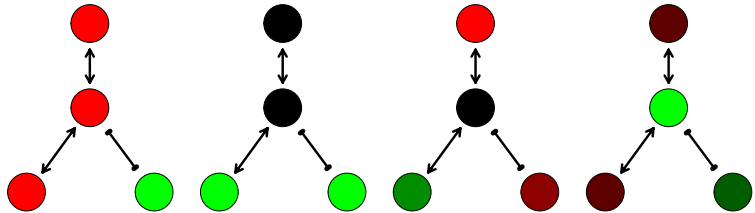}

\caption{Eigenvectors of the signed Laplacian $\LL_{\mathrm{sign}}$
for the simple undirected four-node graph of
example \protect\eqref{eq:adj1}. The eigenvectors of $M_\G$ for this
particular network are the same. The corresponding eigenvalues are
$0,1,1,\frac{16}{3}$ for $M_\G$ and $0, 1, 1, 4$ for
$\LL_{\mathrm{sign}}$. Nodes are colored according to the value of
the eigenvector, where green corresponds to high positive values,
red to high negative values, and black to $0$. ``T''-shaped edges
have negative weights.}
\label{fig:ev}
\end{figure}

Following our principle to build a lower dimension space, we use the
first few eigenvectors of $M_\G$ to obtain orthonormal functions with
low energy. As an example, Figure~\ref{fig:ev} displays the
eigenvectors of $M_\G$ for a simple four-node graph with
%
%
\begin{equation}
\label{eq:adj1}
D =
\pmatrix{
1 & 0 & 0 & 0\vspace*{2pt}\cr
0 & 3 & 0 & 0\vspace*{2pt}\cr
0 & 0 & 1 & 0\vspace*{2pt}\cr
0 & 0 & 0 & 1},\qquad
A =
\pmatrix{
0 & 1 & 0 & 0\vspace*{2pt}\cr
1 & 0 & 1 & -1\vspace*{2pt}\cr
0 & 1 & 0 & 0\vspace*{2pt}\cr
0 & -1 & 0 & 0},
\end{equation}
where $A$ takes on negative values for negative interactions, such as
expression inhibition. The first eigenvector, corresponding to the
smallest energy (eigenvalue of zero), can be viewed as a ``constant''
function on the graph, in the sense that its absolute value is
identical for all nodes, but nodes connected by an edge with negative
weight take on values of opposite sign. By contrast, the last
eigenvector, corresponding to the highest energy, is such that nodes
connected by positive edges take on values of opposite sign and nodes
connected by negative edges take on values of the same sign. Note
that, for this particular example, the adjacency matrix is symmetric,
which need not always be the case. Here, the signed
Laplacian turns out to have the same eigenvectors as $M_\G$, which is not
the case generally.

\begin{figure}

\includegraphics{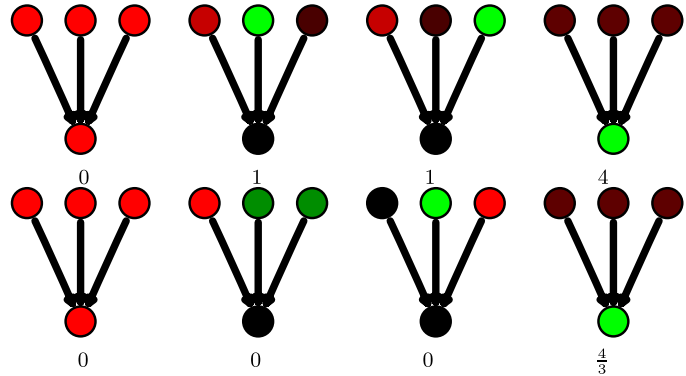}

\caption{Eigenvectors of the signed Laplacian $\LL_{\mathrm{sign}}$
(top) and of $M_\G$ (bottom) for the simple directed four-node
graph of example~\protect\eqref{eq:adj2}. The corresponding eigenvalues
are $0, 1, 1, 4$ and $0, 0, 0, \frac{4}{3}$, respectively. Nodes
are colored according to the value of the eigenvector, where green
corresponds to high positive values, red to high negative values,
and black to~$0$.}
\label{fig:evDir}
\end{figure}

Consider now a slightly different graph, with directed edges, only
positive interactions to avoid confusion, and adjacency matrix
%
%
\begin{equation}
\label{eq:adj2}
A =
\pmatrix{
0 & 1 & 0 & 0\vspace*{2pt}\cr
0 & 0 & 0 & 0\vspace*{2pt}\cr
0 & 1 & 0 & 0\vspace*{2pt}\cr
0 & 1 & 0 & 0}.
\end{equation}
For this graph, Figure~\ref{fig:evDir} shows that the two notions of
energy lead to two different bases. While the signed Laplacian matrix
(by definition based on a~symmetrized version of $A$ for an undirected
graph) has only one (constant) eigenvector of null energy, two of
energy $1$, and one of $4$, $M_\G$ has three orthogonal vectors of
null energy. Note, however, that the first and last eigenvectors are
still the same across the two bases.

More generally, this illustration suggests that projecting on the first
eigenvectors of $M_\G$ will not preserve the same shifts as projecting
on the first eigenvectors of $\LL_{\mathrm{sign}}$. It is possible
for a
shift vector to have low energy~\eqref{eq:energy} but larger signed
Dirichlet energy $\sum_{i,j\in\V}(\delta_i -
a_{ij}\delta_j)^2$, where we recall that~$a_{ij}$ is $1$ for
an edge indicating a positive interaction between $i$ and $j$ and $-1$
for an edge indicating a negative interaction. This is, for example, the
case of the second eigenvector of $M_\G$ on the bottom row of
Figure~\ref{fig:evDir}. It is therefore conceivable that such
a shift essentially lies in the space spanned by the first few
eigenvectors of $M_\G$, but that its projection in the space formed by
the first few eigenvectors of $\LL_{\mathrm{signed}}$ is smaller. As a
consequence, for a particular shift using one basis or the other for
dimensionality reduction will lead to more or less gain in power,
which means that the choice of basis should be adapted to the expected
type of smoothness of the shift.

While we introduce the idea in the context of gene regulation
networks and testing for differential expression, the same
dimensionality reduction principle applies to any multivariate testing
problem for which the variables have a known structure, as represented
by a graph.

As a last remark, we emphasize that our requirement that the shift be
coherent with the network is not too strict in practice. It may sound
like most pairs of nodes must have shifts whose directions are
consistent with the nature of the edge connecting the nodes, but:
\begin{itemize}
\item In practice, keeping a few eigenvectors already allows to
represent several types of shifts which are not perfectly coherent
with the network, as illustrated on Figure~\ref{fig:ev}. The
projection only shrinks those shifts which severely contradict the
prior given by the network.
\item In Section~\ref{sec:synexpErr} we illustrate the fact that this
type of projection still leads to gain in power even in case of
strong misspecifications in the network, {that is,} when a lot of
edges are missing or wrong.
\item\citet{Lopes2011More} show that in a high dimension, \textit{random
projection} of the data in a lower dimension space yields gains in
power against the regular Hotelling $T^2$ in the presence of
correlation and if the spectrum of the covariance matrix decays fast
enough. This result suggests that there is hope to gain power even
in the case where the network doesn't bring much information about
the shift.
\end{itemize}

In the remainder of this paper we denote by $\tilde{f} = U^\top f$
the coefficients of a vector $f\in\RR^{|\V|}$ after projection
on a basis $U$ (typically the eigenvectors of a $Q_\G$ matrix).

\section{Graph-structured two-sample test of means under smooth-shift
alternatives}
\label{sec:test}

For multivariate normal distributions, Hotelling's $T^2$-test, a~classical test of location shift, is known to be a uniformly most
powerful invariant against global-shift alternatives. The test
statistic is based on the squared \textit{Mahalanobis norm} of the
sample mean shift and is given by $T^2 =
\frac{n_1n_2}{n_1+n_2}(\bar{x}_1 - \bar{x}_2)^\top\hat{\Sigma}^{-1}
(\bar{x}_1 - \bar{x}_2)$, where $n_i$, $\bar{x}_i$ and
$\hat{\Sigma}$ denote, respectively, the sample sizes, means and
pooled covariance matrix, for random samples drawn from two
$p$-dimensional Gaussian distributions, $\N(\mu_i,\Sigma)$,
$i=1,2$. Under the null hypothesis $\HZ :\mu_1=\mu_2$ of equal means,
$N T^2$ follows a (central) $F$-distribution $\cF(p,n_1+n_2-p-1)$, where
$N = \frac{n_1+n_2-p-1}{(n_1+n_2-2)p}$. In general, $N T^2$ follows a
noncentral $F$-distribution $\ncF(\frac{n_1
n_2}{n_1+n_2}\Delta^2(\delta,\Sigma); p,n_1+n_2-p-1)$, where the
noncentrality parameter is a function of the Mahalanobis norm of the
mean shift $\delta$, $\Delta^2(\delta,\Sigma) = \delta^\top
\Sigma^{-1}\delta$, which we refer to as the \textit{distribution
shift}. In
the remainder of this paper, unless otherwise specified,
$T^2$-statistics are assumed to follow the nominal $F$-distribution,
{for example}, for critical value and power calculations.

For any orthonormal basis $U$ and, in particular, for our graph-based
basis, direct calculation shows that $T^2\!=\!\tilde{T}^2
\!\stackrel{\Delta}{=}\!\frac{n_1n_2}{n_1+n_2}(\bar{x}_1\!-\!\bar{x}_2)^\top U (U^\top\hat{\Sigma} U)^{-1}U^\top
(\bar{x}_1 - \bar{x}_2)$, {that is}, the statistic $T^2$ in the
original space and the statistic~$\tilde{T}^2$ in the new graph-based
space are identical. More generally, for $k \leq p$, the statistic in
the original space after filtering out dimensions above $k$ is the
same as the statistic $\tilde{T}_k^2$ restricted to the first $k$
coefficients in the new space defined by $U$:\looseness=-1
\begin{eqnarray*}
\tilde{T}_k^2
&\stackrel{\Delta}{=}& \frac{n_1n_2}{n_1+n_2}(\bar{x}_1 -
\bar{x}_2)^\top U_{[k]} \bigl(U_{[k]}^\top\hat{\Sigma}
U_{[k]}\bigr)^{-1}U_{[k]}^\top(\bar{x}_1 - \bar{x}_2) \\
&= &\frac{n_1n_2}{n_1+n_2}(\bar{x}_1 - \bar{x}_2)^\top U 1_{k} U^\top
(U 1_{k} U^\top\hat{\Sigma}
U 1_{k} U^\top)^{+}U 1_{k} U^\top(\bar{x}_1 -
\bar{x}_2),
\end{eqnarray*}\looseness=0
where $A^{+}$ denotes the generalized inverse of a matrix $A$, the $p
\times k$ matrix~$U_{[k]}$ denotes the restriction of $U$ to its first
$k$ columns, and $1_{k}$ is a $p \times p$ diagonal matrix, with $i$th
diagonal element equal to one if $i \leq k$ and zero otherwise. Note
that, as retaining the first $k$ dimensions corresponds to a
\textit{noninvertible} transformation, this filtering indeed has an
effect on the test statistic, that is, we have $\tilde{T}_k^2 \neq
\tilde{T}^2$ in general. As the Mahalanobis norm is invariant to
invertible linear transformations, using an invertible filtering (such
as weighting each component according to its corresponding eigenvalue)
would have no impact on the test statistic.

Hotelling's $T^2$-test is known to behave poorly in the high dimension;
Lemma~$1$ stated and proved in the supplemental
article \hyperref[app:technical]{Supplement A} [\citet{Jacob2011More-SuppA}]
shows that gains in power can be achieved by filtering. Specifically,
let $\tilde{\delta} = U^{\top}\delta$ and $\tilde{\Sigma} =
U^\top\Sigma U$ denote, respectively, the mean shift and covariance
matrix in the new space. Given $k \leq p$, let
$\Delta_{k}^2(\delta,\Sigma) =
\delta_{[k]}^\top(\Sigma_{[k]})^{-1}\delta_{[k]}$ denote
the distribution shift restricted to the first~$k$ dimensions of
$\delta$ and $\Sigma$, {that is}, based on only the first~$k$
elements of $\delta$, $(\delta_i: i\leq k)$, and the first $k \times
k$ diagonal block of $\Sigma$, $(\sigma_{ij}: i, j \leq k)$. Under the
assumption that the distribution shift is smooth, {that is}, lies
mostly in the first few graph-based coefficients, so that
$\Delta_{k}^2(\tilde{\delta},\tilde{\Sigma})$ is nearly maximal
for a~small value of $k$, Lemma~$1$ states that performing
Hotelling's test in the new space restricted to its first $k$
components yields more power than testing in the entire new
space. Equivalently, the test is more powerful in the original space
after filtering than in the original unfiltered space. The increase in
shift $\eta(\alpha,k,l)$ required to maintain power when increasing
dimension can be evaluated numerically for any $(\alpha,k,l)$. Note
that this result holds because retaining the first $k$ new components
is a \textit{noninvertible} transformation.

\begin{figure}

\includegraphics{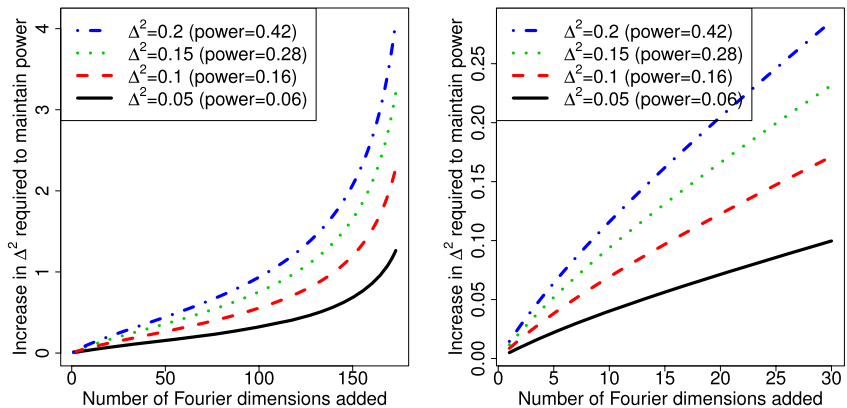}

\caption{Left: Increase in distribution shift required for
Hotelling's $T^2$-test to maintain a given power when increasing
the number of tested new coefficients: $\Delta_{k+l}^2 -
\Delta_{k}^2$ vs. $l$ such that
$\beta_{\alpha,k+l}(\Delta_{k+l}^2)=\beta_{\alpha,k}(\Delta_k^2)$.
Power $\beta_{\alpha,k+l}(\Delta_{k+l}^2)$ computed under the
noncentral $F$-distribution $\ncF(\frac{n_1
n_2}{n_1+n_2}\Delta_{k+l}^2; k+l,n_1+n_2-(k+l)-1)$, for
$n_1= n_2= 100$ observations, $k=5$, and $\alpha=10^{-2}$. Each
line corresponds to the fixed shift $\Delta_{k}^2$ and power
$\beta_{\alpha,k}(\Delta_{k}^2)$ pair indicated in the
legend. Right: Zoom on the first $30$ dimensions.}
\label{fig:shiftInc}
\end{figure}

Corollary~$1$ in the supplemental article \hyperref[app:technical]{Supplement A}
[\citet{Jacob2011More-SuppA}] states that if the distribution shift lies
in the first $k$ new coefficients, then testing in this subspace
yields strictly more power than using additional coefficients. In
particular, if there exists $k<p$ such that $\tilde{\delta}_j = 0$
$\forall j>k$ ({i.e.}, the mean shift is smooth) and
$\tilde{\Sigma}$ is block-diagonal such that $\tilde{\sigma}_{ij} =
0$ $\forall i<k, j>k$, then gains in power are obtained by testing in
the first $k$ new components. Although nonnecessary, this condition
is plausible when the mean shift lies at the beginning of the spectrum
(i.e., has low energy), as the coefficients which do not contain the
shift are not expected to be correlated with the ones that do contain
it.

Note that the result in Lemma~$1$ is more general, as
testing in the first~$k$ new components can increase power even
when the distribution shift partially lies in the remaining
components, as long as the latter portion is below a certain
threshold. Figure~\ref{fig:shiftInc} illustrates, under different
settings, the increase in distribution shift necessary to maintain a
given power level against the number of added coefficients.

Under the assumption of block-diagonal covariance, Corollary~$2$ (in
the supplemental article~\hyperref[app:technical]{Supplement A}
[\citet{Jacob2011More-SuppA}]) directly relates the energy of the mean
shift vector to the gain in power. It states that if the energy of the
mean shift vector $\delta$ is small enough, {that is}, if the mean
shift is coherent enough with the network, then testing in the first
$k$ dimensions of the new basis is more powerful than testing in the
original space. The corresponding upper bound on the mean shift energy
can be quantified for a given generative setting
($\mu_1,\mu_2,\Sigma$), graph $\G$ and level $\alpha$. Tighter and
looser bounds can be straightforwardly derived using the same
principle for the diagonal and general covariance cases, respectively.

\begin{algorithm}
{\fontsize{9.9}{11.9}\selectfont{
  \SetKwData{cSG}{subgraph}
  \SetKwData{ccSG}{q-subgraph}
  \SetKwData{sg}{previousSubgraph}
  \SetKwData{previousSGs}{{\sg}s}
  \SetKwData{checkedSGs}{checkedSubgraphs}
  \SetKwData{prunedSGs}{prunedSubgraphs}
  \SetKwData{SGs}{currentSubgraphs}
  \SetKwData{ub}{$\tilde{T}^2_{k, \max}$}
  \SetKwData{res}{selectedSubgraphs}
  \SetKwFunction{nodes}{nodes}
  \SetKwFunction{boundary}{subgraphBoundary}
  \SetKwFunction{pValue}{p}
  \SetKwFunction{upperBound}{upperBound}
  \SetLine
  \KwIn{ $\G, X_1, X_2, \alpha, q$}
  \KwOut{\res}
  \res $= \varnothing$\;
  \previousSGs = \nodes($\G$)\;
  \prunedSGs $= \varnothing$\;
  \ForEach{$s \in \{1 \dots q-1\}$}{
    \checkedSGs $= \varnothing$\;
    \ForEach{\sg}{
      \ForEach{$\cSG \in \boundary(\sg)$}{
        \lIf{\cSG has been checked or has a pruned subgraph}{next}\;
        \eIf{$s < q-1$}{
          \eIf{$\upperBound(\cSG, \G, X_1, X_2, q) < T^2_{\alpha, k}$}{
            add \cSG to \prunedSGs;
          }{
            add \cSG to \SGs;
          }
        }{
          \ForEach{$\ccSG \in \boundary(\cSG)$}{
            \eIf{\ccSG has been checked or has a pruned subgraph}{next}{
              \If{$\tilde{T}^2_{k}(\ccSG, X_1, X_2) > T^2_{\alpha, k}$}{
                add \ccSG to \res
              }
              add \ccSG to \checkedSGs
            }
          }
        }
        add \cSG to \checkedSGs
      }
    }
    set \previousSGs to \SGs
  }}}
  \caption{Nonhomogeneous subgraph discovery algorithm. The~\texttt{sub\-graphBoundary} of a subgraph
  $g$ of $\G$ is defined as the set of supergraphs of $g$ obtained by adding any one node
  of $\G$ which is connected to a node of~$g$.}
  \label{algo:exact}
\end{algorithm}

\section{Nonhomogeneous subgraph discovery}
\label{sec:discovery}

A systematic approach for discovering nonhomogeneous subgraphs,
{that is}, subgraphs of a large graph that exhibit a significant
shift in means, is to test all of them one-by-one.

This poses a huge combinatorial problem even for moderately large
graphs ($p=50$, say), as the number of (connected) subgraphs of size
$k$ of a graph of size $p$ can be exponential in $p$ and $k$.
Exhaustive search is therefore not feasible in practice, especially
for differential expression on gene networks, where~$p$ is typically
in the dozens or hundreds of genes. Therefore, it is important to
rapidly identify sets of subgraphs that all satisfy the null
hypothesis of equal means. To this end, we prove an upper bound on
the value of the test statistic for any subgraph containing a given
set of nodes (Lemma~$2$ in the supplemental
article~\hyperref[app:technical]{Supplement A} [\citet{Jacob2011More-SuppA}]). An exact
algorithm is derived from this upper bound in Section~\ref{sec:exact},
and a~quicker, approximate algorithm is proposed in
Section~\ref{sec:euclidean}.


\subsection{Exact algorithm}
\label{sec:exact}
Given a large graph $\G$ with $p$ nodes, we adopt a~branch-and-bound-like approach [\citet{Land1960Automatic}] to test
subgraphs of size $q \leq p$ at level $\alpha$, as described in
Algorithm~\ref{algo:exact}. We start by checking, for each node in~$\G$, whether the Hotelling $T^2$-statistic in the first $k$ new
components of any subgraph of size $q$ containing this node can be
guaranteed (by virtue of Lemma~$2$) to be below the level-$\alpha$
critical value $T^2_{\alpha,k}$, {for example}, $(1-\alpha
)$-quantile of
$\cF(k,n_1+n_2-k-1)$ distribution. If this is the case, the node is
pruned, that is, removed from the graph. The algorithm iteratively
enriches a list of pruned subgraphs and a list of candidate subgraphs
(called \texttt{prunedSubgraphs} and \texttt{currentSubgraphs} in
Algorithm~\ref{algo:exact}, resp.) of increasing number of
nodes $s=1,\ldots, q-1$. Pruned subgraphs are those for which one can
guarantee that no supergraph of size $q$ can reach significance level
$\alpha$, and candidate subgraphs are those for which this guarantee
cannot be given. The key of the algorithm is that at step $s$, only
those graphs containing a~candidate subgraph have to be considered.

This guarantee is obtained by applying Lemma~$2$ in the supplemental
article~\hyperref[app:technical]{Supplement A} [\citet{Jacob2011More-SuppA}], which gives
an upper bound on the value of the test statistic for any subgraph
containing a~given set of nodes. For a subgraph $g$ of $\G$ of size
$q \leq p$, Hotelling's $T^2$-statistic in the first $k \leq q$ new
components of $g$ is defined as
\[
\tilde{T}_k^2(g)=\frac{n_1 n_2}{n_1+n_2}\bigl(\bar{x}_1(g) -
\bar{x}_2(g)\bigr)^\top U_{[k]} \bigl( U_{[k]}^\top\hat{\Sigma}(g) U_{[k]}
\bigr)^{-1}U_{[k]}^\top\bigl(\bar{x}_1(g) - \bar{x}_2(g)\bigr),
\]
where $U_{[k]}$ is the $q \times k$ restriction of the matrix of $q$
eigenvectors of $Q_g$ to its first $k$ columns [{i.e.},
$U_{[k]}(g)$, where we omit $g$ to ease notation] and $\bar{x}_i(g),
i=1,2$, and $\hat{\Sigma}(g)$ are, respectively, the empirical means
and pooled covariance matrix restricted to the nodes in $g$.
Lemma~$2$ states that for any number $k$ of retained components, and
for any subgraph $g'$ of size $q'$ of $g$, $\tilde{T}_k^2(g)$ is upper
bounded by the~$T^2$ statistic of the subgraph whose nodes are in $\nu
(g', q-q')$, that is, the union of the nodes of $g'$ and the nodes of
$g$ whose shortest path to a node of $g'$ is less than or equal to $r$.
The set $\nu(g',r)$ is called the $r$-neighborhood of $g'$.
As a corollary of Lemma~$2$, the subgraphs returned by Algorithm~\ref
{algo:exact} are exactly those who exhibit a significant shift in means
at the prescribed level $\alpha$.

Note that the bound in Lemma~$2$ takes into account the fact that the
$T^2$-statistic is eventually computed in the first few components of
a basis which is not known beforehand: at each step, for each
potential subgraph~$g'$ which would include the subgraph $g$ which we
consider for pruning, the $\tilde{T}_k^2(g')$ that needs to be bounded
above depends on $Q_{g'}$.

\subsection{Mean-shift approximation}
\label{sec:euclidean}

For ``small-world'' graphs above a certain level of connectivity and
$q$ large enough, $\nu(g',q-s)$, the $(q-s)$-neighbor\-hood of $g'$,
tends to be large, at least at the beginning of the above exact
algorithm, and the number of tests actually performed may not decrease
much compared to the total number of possible tests. One can, however,
identify much more efficiently the subgraphs whose sample mean shift
in the first $k$ components of the new space has Euclidean
norm $\|\hat{\tilde{\delta}}_{[k]}(g)\| = \|U_{[k]}^\top(\bar{x}_1(g)
- \bar{x}_2(g))\|$ above\vadjust{\goodbreak} a certain threshold. Indeed, it is
straightforward to see that
\begin{eqnarray*}
&&\bigl\|U_{[k]}^\top\bigl(\bar{x}_1(g) - \bar{x}_2(g)\bigr)\bigr\|^2\\
&&\quad\leq
\bigl\|U^\top\bigl(\bar{x}_1(g) - \bar{x}_2(g)\bigr)\bigr\|^2\\
&&\quad= \|\bar{x}_1(g) - \bar{x}_2(g)\|^2 \\
&&\quad\leq\|\bar{x}_1(g')-\bar{x}_2(g')\|^2 \\
&&\qquad{}+  \max_{v_1,\ldots,v_{q-s}\in\nu(g',q-s)}\|\bar{x}_1(v_1,\ldots
,v_{q-s})-\bar{x}_2(v_1,\ldots,v_{q-s})\|^2.
\end{eqnarray*}
Using this inequality yields an upper bound on $\tilde{T}_k^2(g)$ that
can be used as \texttt{upperBound} at line 10 of
Algorithm~\ref{algo:exact}. This defines a procedure that identifies
all subgraphs for which the Euclidean norm of the sample mean shift
exceeds a given threshold: $\|\hat{\tilde{\delta}}_{[k]}(g)\|^2 >
\theta$. For any $\alpha$, if this threshold $\theta$ is low enough,
all the subgraphs with $\tilde{T}_k^2(g)>T^2_{\alpha,k}$ are included
in this set. Performing the actual $T^2$-test on these preselected
subgraphs then yields exactly the set of subgraphs that would have
been identified using the exact procedure of
Section~\ref{sec:exact}. More precisely, Lemma~$4$, in the
supplemental article \hyperref[app:technical]{Supplement A} [\citet{Jacob2011More-SuppA}],
states that for any subgraph which would be detected by Hotelling's
$T^2$-statistic $\tilde{T}_k^2(g)$ but not by the Euclidean criterion
$\|\hat{\tilde{\delta}}_{[k]}(g)\|^2$, the sample covariance matrix in
the restricted new space (after filtering) has an eigenvalue below a
certain threshold. This implies that such false negative subgraphs
(from the Euclidean approximation to the exact algorithm) have a small
mean shift in the new space, but in a direction of small variance. In
the context of gene expression, this is related to the well-known issue of
the detection of DE genes by virtue of their small variances. Even
though the differences in expression appear to be significant for
these genes, they correspond to small effects that may not be
interesting from a practical point of view ({i.e.}, biologically
nonsignificant). Methods for addressing this problem are proposed in
\citet{Lonnstedt2001Replicated}.\vspace*{-1.5pt} Note that
$\lambda_{\min}(\hat{\Sigma}(g)) \leq
\lambda_{\min}(\hat{\tilde{\Sigma}}_{[k]}(g))$; thus, the remark on
variances holds for both the new and the original spaces. However, if
$q$ is large, we expect $\lambda_{\min}(\hat{\Sigma}(g))$ to be very
small, while filtering somehow controls the conditioning of the
covariance matrix.

\subsection{Multiple hypothesis testing}
\label{sec:multiple-testing}

Testing for homogeneity over the potentially large number of subgraphs
investigated as part of the above algorithms immediately raises the
issue of multiple testing. However, because one does not know in
advance the total number of tests and which tests will be performed
specifically, standard multiple testing procedures, such as those in
\citet{Dudoit2008Multiple}, are not immediately applicable.

In an attempt to address the multiplicity issue, we apply a
permutation procedure to control the number of false positive
subgraphs under\vadjust{\goodbreak} the complete null hypothesis of identical
distributions in the two populations. Specifically, one permutes the
class/population labels (1 or 2) of the $n_1+n_2$ observations and
applies the nonhomogeneous subgraph discovery algorithm to the
permuted data to yield a certain number of false positive
subgraphs. Repeating this procedure a sufficiently large number of
times produces an estimate of the distribution of the number of Type I
errors under the complete null hypothesis of identical distributions.

We evaluate the empirical behavior of the procedures proposed in
Sections~\ref{sec:test} and \ref{sec:discovery}, first on synthetic
data, then on breast cancer microarray data analyzed in the context of
KEGG pathways.

\section{Results on synthetic data}
\label{sec:synexp}

The performance of the graph-structured test is assessed in cases
where the distribution shift $\Delta^2$ satisfies the smoothness
assumptions described in Section~\ref{sec:test}. We first generate a
connected random graph $\G$ with $p=20$ nodes and $20$ edges. Next,
we generate $10\mbox{,}000$ data sets in the space corresponding to the basis
$U$ defined by the eigenvectors of the $Q_\G$ matrix for the graph
$\G$; an inverse transformation is applied to random vectors generated
is this new space. Each data set comprises $n_1=n_2=20$ Gaussian random
vectors in $\RR^{p}$, with null mean shift $\delta$ for $5000$
data sets and nonnull mean shift $\delta$ for the remaining
$5000$. For the latter data sets, the nonzero shift is built in the
first $k_0=3$ graph-based coefficients (the shift being zero for the
remaining $p-k_0$ coefficients): $\tilde{\delta}_i \neq0$ if and only
if $i \leq k_0$ and $\Delta^2(\delta,\Sigma) =
\Delta^2(\tilde{\delta},\tilde{\Sigma}) =
\tilde{\delta}^\top\tilde{\Sigma}^{-1}\tilde{\delta}=1$. We consider
two covariance settings. In the first one, the covariance matrix in
the new space is diagonal, with diagonal elements equal to
$\frac{1}{\sqrt{p}}$. In the second, correlation is introduced between
the shifted coefficients only. Specifically, for $i,j \leq k_0$,
$\tilde{\Sigma}_{ij}=\frac{0.5}{\sqrt{p}}$ if $i\neq j$,
$\tilde{\Sigma}_{ii}=\frac{0.9}{\sqrt{p}}$ otherwise.

\subsection{Fixed known network}
\label{sec:synexpclean}

\begin{figure}

\includegraphics{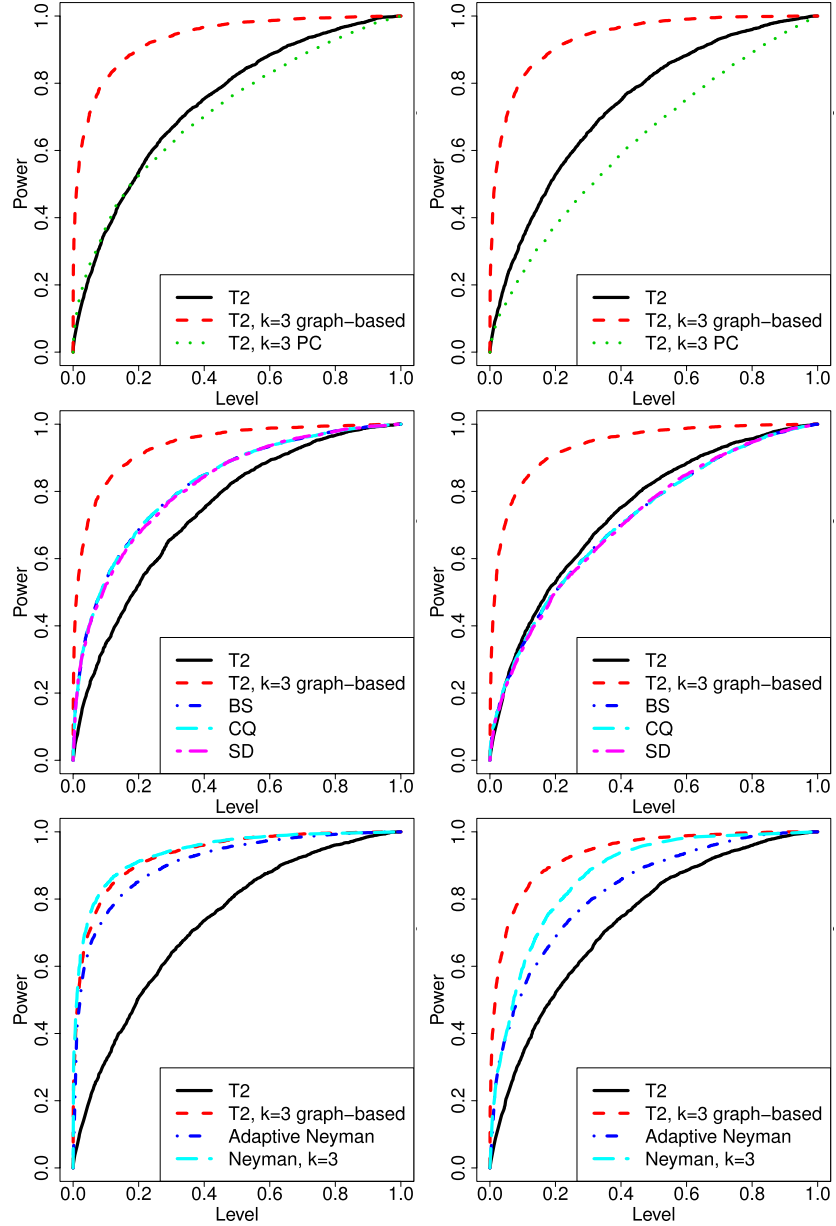}

\caption{{Synthetic data: ROC curves for the detection of a
smooth shift.} Left: Diagonal covariance structure. Right:
Block-diagonal covariance structure. Top: Comparison of tests
based on the standard Hotelling $T^2$-statistic in the original
space, $T^2$-statistic in the first $k_0$ graph-based
coefficients, and $T^2$-statistic in the first $k_0$ principal
components. Middle: Comparison with the statistics
of~\textit{Bai and Saranadasa} (\protect\citeyear{Bai1996Effect}) (BS), \textit{Chen and Qin} (\protect\citeyear{Chen2010A}) (CQ), and
\textit{Srivastava and Du} (\protect\citeyear{Srivastava2008Test}) (SD). Bottom: Comparison with the
Neyman statistics of \textit{Fan and Lin} (\protect\citeyear{Fan1998Test}).}
\label{fig:rocs}
\end{figure}

Figure~\ref{fig:rocs} displays receiver operator characteristic (ROC)
curves for mean shift detection by the standard Hotelling $T^2$-test,
$T^2$ in the first $k_0$ graph-based coefficients, $T^2$ in the first $k_0$
principal components (PC), the adaptive Neyman test of
\citet{Fan1998Test}, and a modified version of this fourth test where
the correct value of $k_0$ is specified. Note that we do not consider
sparse learning
approaches [\citet{Jacob2009Group}, \citet{Jenatton2009Structured}, \citet{Obozinski2011Group}], but it would
be straightforward to design a realistic setting where such approaches
are outperformed by testing, {for example}, by adding correlation
between some of the functions under $\HO$.

\begin{figure}

\includegraphics{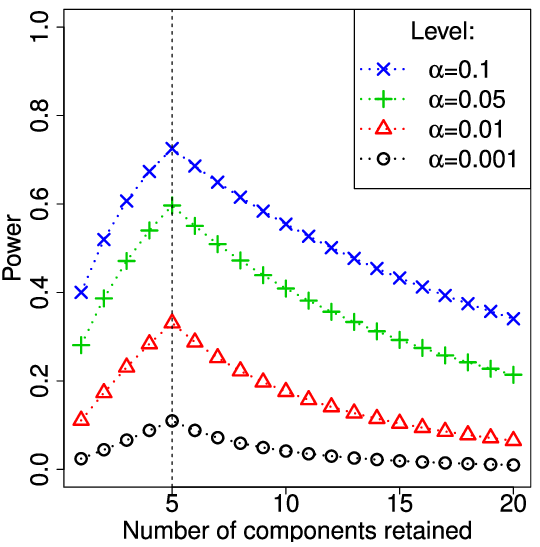}

\caption{{Synthetic data: Sensitivity to choice of $k$.} Power
of the $T^2$-test in the first $k$ graph-based coefficients for a
graph of $20$ nodes, when the actual distribution shift $\Delta^2
= 1$ is evenly distributed among the first $k_0=5$ graph-based
coefficients and with $n_1=n_2=20$.}
\label{fig:miss}
\end{figure}

The first important comparison is between the classical Hotelling
$T^2$-test vs. the $T^2$-test in the new graph-based space (Figure~\ref
{fig:rocs}, top row). As expected from
Lemma~$1$, testing in the restricted space where the shift
lies performs much better than testing in the full space which
includes irrelevant coefficients. The difference can be made
arbitrarily large by increasing the dimension $p$ and keeping the
shift unchanged. The graph-structured test retains a large advantage
even for moderately smooth shifts, {for example}, when $k_0=3$ and
$p=5$. Of course, this corresponds to the optimistic case where the
number of shifted coefficients $k_0$ is known. Figure~\ref{fig:miss}
shows the power of the test in the new space for various choices of
$k$. Even when missing some coefficients ($k<k_0$) or adding a few
irrelevant ones ($k>k_0$), the power of the graph-structured test is
higher than that of the $T^2$-test in the full space. The principal
component approach is shown in Figure~\ref{fig:rocs} (top row) because
it was proposed for the
application which motivated our work [\citet{Ma2009Identification}] and
because it also illustrates that the improvement in performance
originates not only from dimensionality reduction, but also from the
fact that this reduction is in a direction that does not decrease the
shift. We emphasize that power entirely depends on the nature of the
shift and that a PC-based test would outperform our graph-based test
when the shift lies in the first principal components rather than
graph-based coefficients.\looseness=-1

The panels in the middle row show that the statistics
of~\citet{Bai1996Effect}, \citet{Chen2010A} and
\citet{Srivastava2008Test} are also largely outperformed by our
graph-structured statistic. This observation suggests that when such a
graph-based prior on the shift is available, working in the new,
lower-dimensional space does better at solving the problem of
high-dimensionality than methods based on diagonal approximations of
the covariance matrix. In addition, as one could expect, the
procedures of~\citet{Bai1996Effect}, \citet{Chen2010A} and
\citet{Srivastava2008Test} perform very poorly in the presence of
correlation. Here again, for a nonsmooth shift,\vadjust{\goodbreak} the comparison would
be less favorable to our procedure. We also considered the
recently-proposed random projection approach of
\citet{Lopes2011More}. Random projection was shown to give more power
than ~\citet{Bai1996Effect}, \citet{Chen2010A} and
\citet{Srivastava2008Test} in high-dimensional cases. However, as
expected in our simulation setting where the sample size is twice the
number of dimensions, it did not improve upon the Hotelling $T^2$-test
(ROC curve not shown for the sake of readability). The method of
\citet{Lopes2011More} is more appropriate in a higher dimension and when
no prior on the shift direction is available.

Finally, we consider the adaptive Neyman test of \citet{Fan1998Test}
(bottom two panels of Figure~\ref{fig:rocs}), which takes advantage of
smoothness assumptions for time series. This test differs from our
graph-structured test, as Fourier coefficients for stationary
time series are known to be asymptotically independent and
Gaussian. For graphs, the asymptotics would be in the number of nodes,
which is typically small, and necessary conditions such as
stationarity are more difficult to define and unlikely to hold for
data such as gene expression measures. In the uncorrelated setting,
the modified version of the \citet{Fan1998Test} statistic based on the
true number of nonzero coefficients performs approximately as well as
the graph-structured $T^2$. However, for correlated data, it loses
power and both versions of the Neyman test can have arbitrarily
degraded performance. This, together with the need to use the
bootstrap to calibrate the test, illustrates that direct transposition
of the \citet{Fan1998Test} test to the graph context is not optimal.

\begin{figure}

\includegraphics{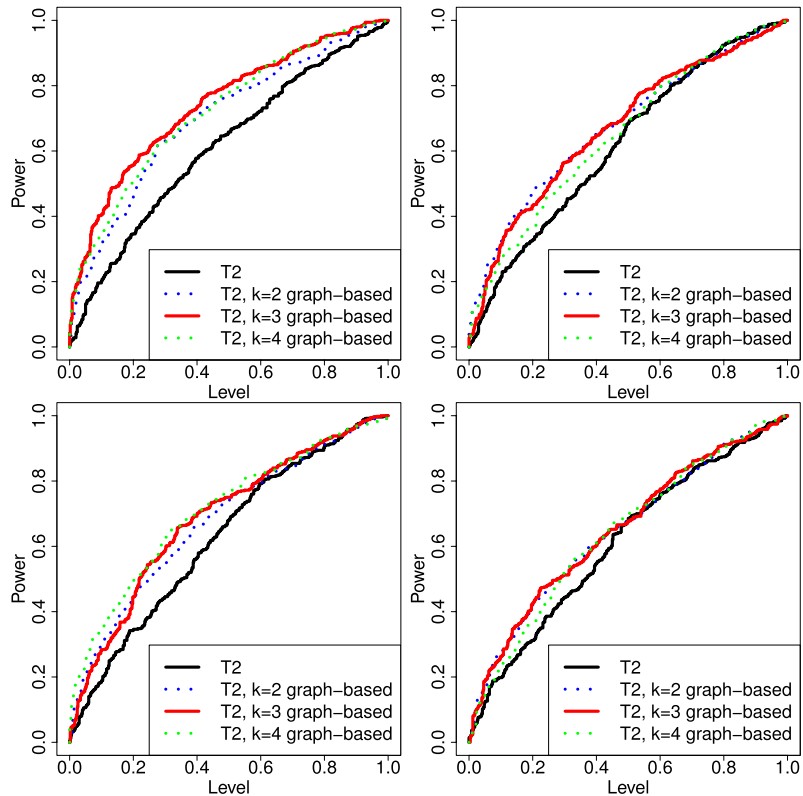}

\caption{{Synthetic data: ROC curves for the detection of a smooth
shift in the presence of errors in the
network.} Comparison of tests based on the standard Hotelling
$T^2$-statistic in the original space and $T^2$-statistic in the
first $k$ graph-based coefficients. Top: After randomly removing
$20$ (left) and $40$ (right) edges. Bottom: After randomly adding
$20$ (left) and $40$ (right) edges.}
\label{fig:toyCorrupt}
\end{figure}

\subsection{Fixed network with errors}
\label{sec:synexpErr}

We now consider the less idealistic case where the network used for
testing is not exactly the one which was used to generate the
data. More precisely, we follow the same procedure as in the
correlated case of Section~\ref{sec:synexpclean}, but remove or add
some edges to the network between the moment where we use it to
generate the two samples and the moment where we use it in our testing
procedure. This setting is much closer to what is likely to happen
with real data, as available networks may miss several gene
interactions which are not known yet and may include some incorrect
interactions or some which are irrelevant for the problem under
consideration. It is easy to see that in a worst case scenario,
removing or adding an edge to the network can arbitrarily shrink the
$T^2_k$ statistic. Take, for example, two disconnected nodes and assume
without loss of generality that the empirical covariance matrix is the
identity matrix and the empirical mean shifts for the two nodes are
$1$ and $-1$. Then, $\delta^\top\delta$ is $2$, but adding an edge
between the two nodes and projecting on the first eigenvector of the
graph Laplacian matrix shrinks the observed shift to $0$. A
probabilistic analysis over random perturbations would be out of the
scope of this paper, but the following simulation study is intended to
give insight into what would happen in practice if randomly chosen
edges are either wrongly added or omitted.\vadjust{\goodbreak}

\begin{figure}

\includegraphics{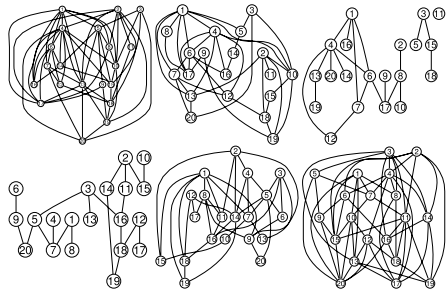}

\caption{{Synthetic data: Examples of corrupted networks used
to generate
Figure~\protect\ref{fig:toyCorrupt}.} Left column: original network used
to generate the data of Section~\protect\ref{sec:synexpErr} before removing (top
row) and adding (bottom row) edges. Middle column: one instance
of removing/adding $20$ edges. Right column: one instance
of removing/adding $40$ edges.}
\label{fig:toyCorruptG}\vspace*{-3pt}
\end{figure}

For the sake of clarity, Figure~\ref{fig:toyCorrupt} only shows ROC
curves for our graph-structured $T^2$ with $k=2,3,4$, and the standard
Hotelling $T^2$-statistic. The other
competitors [\citet{Bai1996Effect},
\citet{Ma2009Identification},
\citet{Chen2010A},
\citet{Fan1998Test},
\citet{Srivastava2008Test},
\citet{Lopes2011More}]
considered above all perform similarly to the Hotelling
$T^2$-statistic. In the case where edges are erroneously removed, we
start with a true network having $60$ edges instead of $20$ in
Section~\ref{sec:synexp}. Figure~\ref{fig:toyCorrupt} shows that our
graph-based approach can still perform much better than all competing
methods, even in cases where the topology of the observed network is
very incomplete ($1/3$ of the true number of edges) or noised by a lot
of spurious edges. Figure~\ref{fig:toyCorruptG} shows examples of
networks corrupted by removing (top row) or adding (bottom row) edges
to the original one and used in this experiment. It is visually clear
that the information provided to our procedure is very different from
the one, that is, used to generate the data. Again, this is an
encouraging result, as it is well known that the gene networks
available in the literature are missing a lot of interactions
and often include incorrect information.

\subsection{Branch-and-bound subgraph discovery}

\begin{figure}

\includegraphics{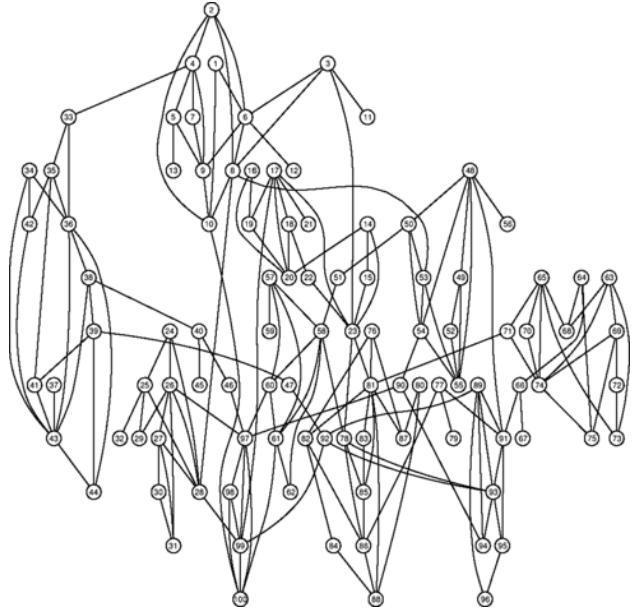}

\caption{{Synthetic data: Random graph used in the evaluation of
the pruning procedure.}}
\label{fig:randg}
\end{figure}

To evaluate the performance of the subgraph discovery algorithms
proposed in Section~\ref{sec:discovery}, we generated a graph of $100$
nodes formed by tightly-connected hubs of sizes sampled from a Poisson
distribution with parameter 10 and only weak connections between these
hubs (Figure~\ref{fig:randg}). Such a graph structure is intended to
mimic the typical topology of gene regulation networks. We randomly
selected one subgraph of $5$ nodes to be nonhomogeneous, with smooth
shift in the first $k_0=3$ coefficients. The mean shift was set to
zero on the rest of the graph. We set the norm of the mean shift to~$1$ and
the covariance matrix to identity, so that detecting the
shifted subgraph is impossible by just looking at the mean shift on
the graph.

We evaluated run-time for full enumeration, the exact branch-and-bound
algorithm based on Lemma~$2$ (Section
\ref{sec:exact}), and the approximate algorithm based on the Euclidean
norm (Section~\ref{sec:euclidean}). We also examined run-time on data
with permuted class labels, as the subgraph discovery procedure is to
be run on such data to evaluate the number of false positives and
adjust for multiple testing. Averaging over $20$ runs, the full
enumeration procedure took $732\pm9$ seconds per run and the exact
branch-and-bound $627\pm59$ seconds on the nonpermuted data and
$578\pm100$ seconds on permuted data. Over $100$ runs, the
approximation at $\theta=0.5$ ($\lambda_{\min}=0.52$) took $204\pm86$
seconds ($129\pm91$ on permuted data) and the approximation at
$\theta=1$ ($\lambda_{\min}=1.04$) took $183\pm106$ seconds ($40\pm
60$ on permuted data). The latter approximation missed the
nonhomogeneous subgraph in $5\%$ of the runs.

While neither the exact nor the approximate bounds are efficient
enough to allow systematic testing on\vadjust{\goodbreak} huge graphs for which the full
enumeration approach would be impossible, they allow a significant
gain in speed, especially for permuted data, and will thus prove to be
very useful for multiple testing adjustment.

\section{Results on cancer gene expression data}
\label{sec:geneexp}

We also validated our methods using the following two microarray
expression data sets: a breast cancer data set [\citet
{Loi2008Predicting}] and a bladder cancer data set [\citet
{Stransky2006Regional}].

\textit{Breast cancer data set}.
The first data set by \citet{Loi2008Predicting} comprises the expression
measures of
$15\mbox{,}737$ genes for $255$ ER$+$ breast cancer patients treated with
tamoxifen. Breast tumors are generally classified into three main
categories [\citet{Perou2000Molecular}]: \textit{luminal epithelial/ER$+$},
\textit{HER2$+$}, and \textit{triple negative}. ER$+$ tumors typically
express estrogen receptors at a high level and often rely on estrogen
for their growth. Tamoxifen is an antagonist of estrogen receptors and
therefore prevents its activation by endogenous estrogen. Some ER$+$
tumors, however, keep growing after being treated with tamoxifen. An
important goal is to detect structured groups of genes which are
differentially expressed between resistant and sensitive patients, as
detecting such groups could help understand resistance mechanisms and
eventually improve ER$+$ breast tumor treatment. Using distant
metastasis-free survival as a primary endpoint, $68$ patients from
this data set are labeled as resistant to tamoxifen and $187$ are
labeled as sensitive to tamoxifen.

\textit{Bladder cancer data set}.
The second data set by
\citet{Stransky2006Regional} consists of the expression measures of
$8323$ genes for $57$ urothelial tumors. Urothelial tumors are
known to be arising and evolving through two distinct pathways, one
typically leading to low-grade noninvasive tumors (Ta tumors), the
other involving more aggressive
tumors [\citet{Bakkar2003FGFR3}, \citet{Knowles2006Molecular}]. These two
subtypes, however, are not distinguishable from simple markers such
as estrogen receptor or HER2 status for breast tumors. Mutation of
the FGFR3 gene is sometimes used as a proxy, as about $70\%$ of the
noninvasive tumors carry it. As this information was unfortunately
not available for this data set, we used the second best proxy which
is tumor stage. We defined two groups: $25$ tumors either at the Ta
or T1 stages (TaT1 group) and $32$ tumors at the T2, T3 or T4
stages (T2$+$ group). The muscle invasive T2$+$ tumors are aggressive
and present a high risk of metastasis, while the Ta tumors have high
recurrence level but low chance of progression into muscle invasive
tumors. Identifying pathways which differ in expression between the
two subtypes could help understand the disease better and improve
its treatment.

\begin{figure}

\includegraphics{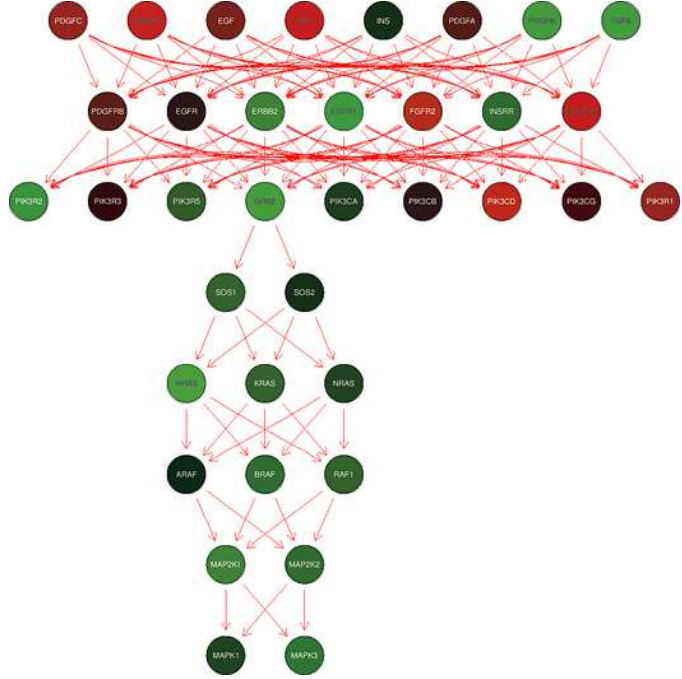}\vspace*{-4pt}

\caption{{Breast cancer data set: KEGG prostate cancer pathway.}
Scaled difference in sample mean expression measures between
tamoxifen-resistant and sensitive patients, for genes in one
component of the KEGG prostate cancer pathway. Nodes are colored
according to the value of the difference in means, with green
corresponding to high positive values, red to high negative
values, and black to $0$. Red arrows denote activation, blue arrows inhibition.}
\label{fig:prostate}\vspace*{-5pt}
\end{figure}

\begin{figure}

\includegraphics{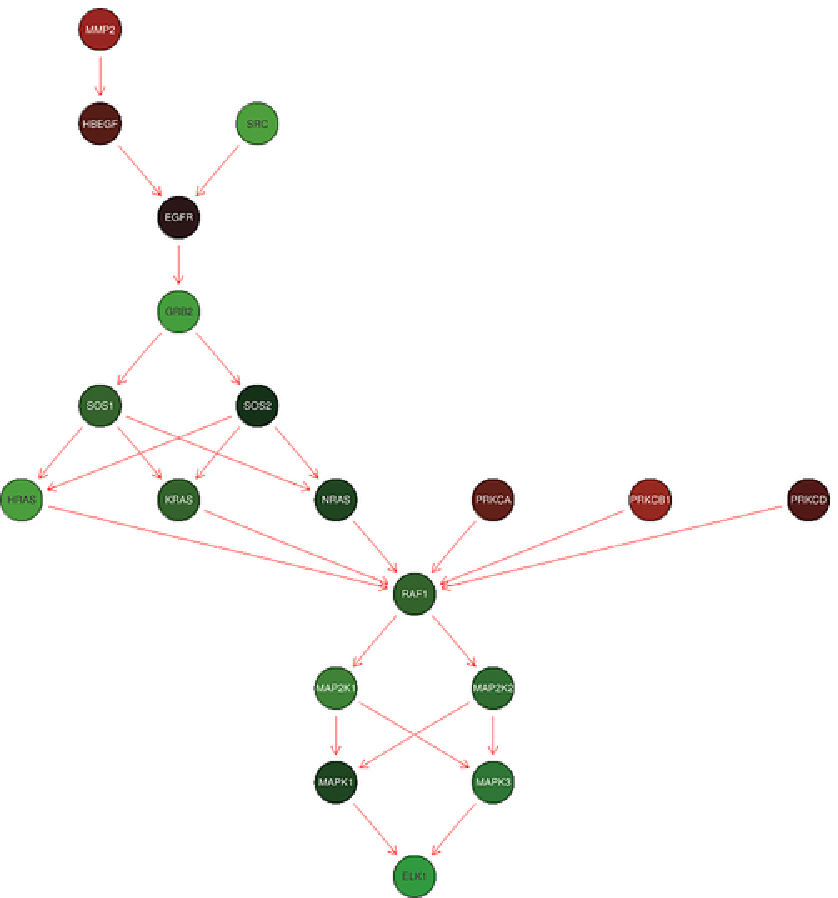}

\caption{{Breast cancer data set: KEGG GnRH signaling
pathway.} Scaled
difference in sample mean expression measures between
tamoxifen-resistant and sensitive patients, for genes in one
component of the KEGG GnRH signaling pathway. Nodes are colored
according to the value of the difference in means, with green
corresponding to high positive values, red to high negative
values, and black to $0$. Red arrows denote activation, blue arrows inhibition.}
\label{fig:gnrh}
\end{figure}

\subsection{KEGG networks}
\mbox{}

\textit{Breast cancer data set}.
Starting with the breast cancer data set, we tested individually $351$
connected components from $100$ KEGG pathways corresponding to known
gene regulation networks listed in the supplemental
article~\hyperref[app:pathwaytables]{Supplement~B} [\citet{Jacob2011More-SuppB}], using the
classical Hotelling $T^2$-test and the $T^2$-test in the new
graph-based space retaining only the first $20\%$ coefficients ($k=0.2
p$). This value was the one chosen
in~\citet{Rapaport2007Classification} on the same networks, accordingly
with an argument based on loss minimization, not on hypothesis
testing. The analysis of~\citet{Lopes2011More} suggests that the
projection method (on random subspaces in their case) is quite robust
to the choice of~$k$. More refined heuristics could be based on
eigengaps, {that is}, on the distances between successive
eigenvalues. Indeed, matrix perturbation results suggest that
eigenspaces can vary a lot even under small perturbations of the
network if the largest discarded eigenvalue is close to the smallest
kept eigenvalue [\citet{Davis1969Some}, \citet{Stewart1990Matrix},
\citet{Ipsen2010Eigenproblem}]. Values of $k$ such that $\lambda_{k} -
\lambda_{k+1}$ is as large as possible could therefore be generally
preferable.

The networks had $36$ nodes in average, with a median of $23$. For
each of the $351$ graphs, (unadjusted) $p$-values were computed under
the nominal $F$-distributions $\cF(p, n_1+n_2-p-1)$ and $\cF(k,
n_1+n_2-k-1)$, respectively. The \citet{Benjamini1995Controlling}
procedure was then applied to control the false discovery rate (FDR)
at level $0.05$.

Since there is no gold standard regarding which pathways are actually
involved in endocrine resistance, practical validation of the entire
set of detected\vadjust{\goodbreak} pathways requires advanced biological expertise and
further experiments and is the subject of ongoing
collaborations. Nonetheless, inspection of our list reveals several
pathways which would not have been detected (or would have been
further down in the list) without accounting for the network structure
and which have recently been shown to be central in tamoxifen
resistance. Many of these pathways involve the Ras/Raf-1/MAPK cascade
[\citet{McGlynn2009RasRaf1MAPK}], like one of the connected
components of the \textit{prostate cancer} pathway shown in
Figure~\ref{fig:prostate} and one connected component of the
\textit{GnRH} pathway shown in Figure~\ref{fig:gnrh}. The former also
involves the overexpressed FGFR1, whose amplification was very
recently shown to be implicated in endocrine therapy resistance by
\citet{Turner2010FGFR1}. The latter pathway involves overexpressed SRC,
which is also a well-studied target when trying to prevent tamoxifen
resistance [\citet{Herynk2006Cooperative}]. Both pathways have a much
smaller $p$-value when accounting for their graph structure than when
testing in the original gene space: $10^{-4}$ vs. $0.02$ for the
prostate cancer pathway and $10^{-3}$ vs. $0.11$ for the GnRH
signaling pathway. This is because the differences in expression of
individual genes are\vadjust{\goodbreak} insufficient to be significant in $36$ and $19$
dimensions, respectively, while the expression shift projected in the
first $8$ and $4$ graph-based directions, respectively, is
significant. Note that the corresponding $p$-values for the
hypergeometric enrichment test are $0.15$ and $0.31$. The complete
gene lists of the two components are reported in Tables~$3$ and~$4$,
respectively, in the supplemental article~\hyperref[app:glist]{Supplement C}
[\citet{Jacob2011More-SuppC}]. Using a system-based approach like our
proposed graph-based test therefore allows us to recover several known
results (which may not have been obvious from the same data when
looking at each gene individually) and may give insight regarding
other resistance mechanisms by highlighting connections between these
results.

\begin{figure}

\includegraphics{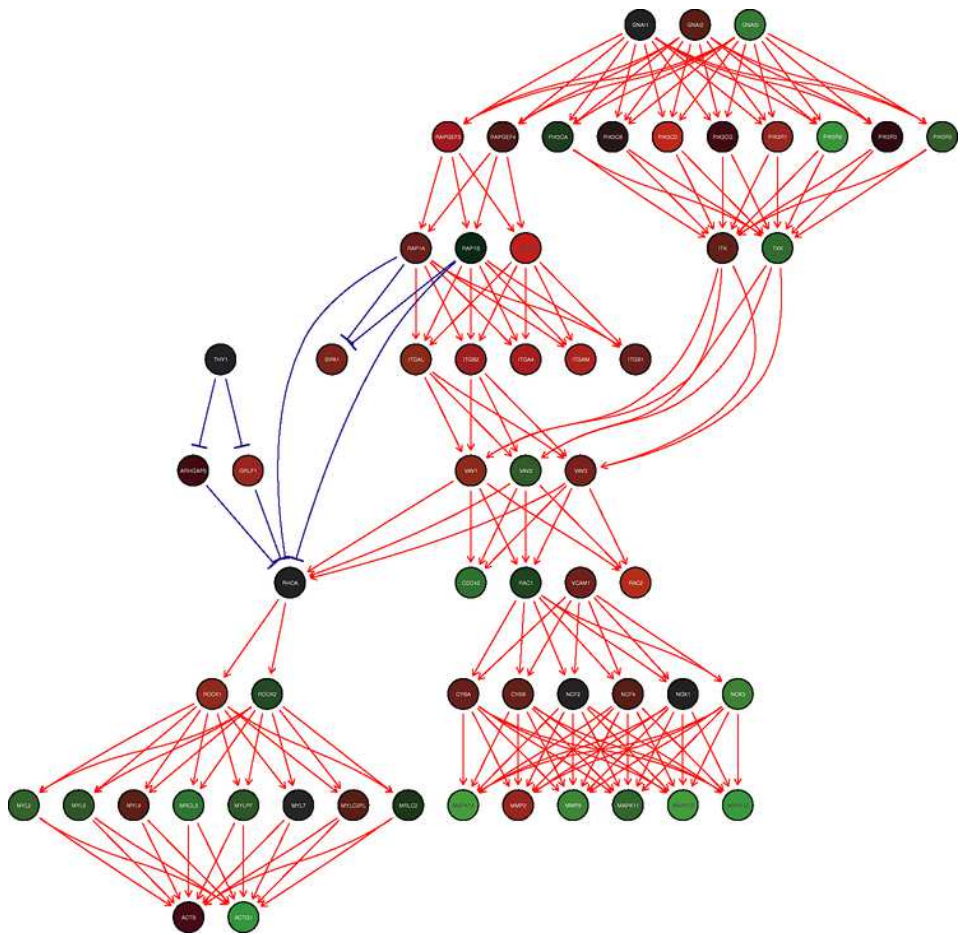}%
\vspace*{-3pt}
\caption{{Breast cancer data set: KEGG leukocyte transendothelial
migration pathway.} Scaled difference in sample mean expression
measures between tamoxifen-resistant and sensitive patients, for
genes in one component of the KEGG leukocyte transendothelial
migration pathway. Nodes are colored according to the value of the
difference in means, with green corresponding to high positive
values, red to high negative values, and black to~$0$. Red arrows
denote activation, blue arrows inhibition.}
\label{fig:leukocyte}\vspace*{-3pt}
\end{figure}

Another example of a network selected only when accounting for graph
structure is \textit{Leukocyte transendothelial migration}, shown in
Figure~\ref{fig:leukocyte}. To the best of our knowledge, this pathway
is not specifically known to be involved in tamoxifen
resistance. However, its role in resistance is plausible, as leukocyte
infiltration was recently found to be involved in breast tumor
invasion [\citet{Man2010Aberrant}]; more generally, the immune system
and inflammatory response are closely related to the evolution of
cancer. Here again, the $p$-value of the hypergeometric test is
extremely high ($0.31$). The entire list of genes in this component is
reported in Table~$5$ in the supplemental article~\hyperref[app:glist]{Supplement C}
[\citet{Jacob2011More-SuppC}].

\begin{figure}

\includegraphics{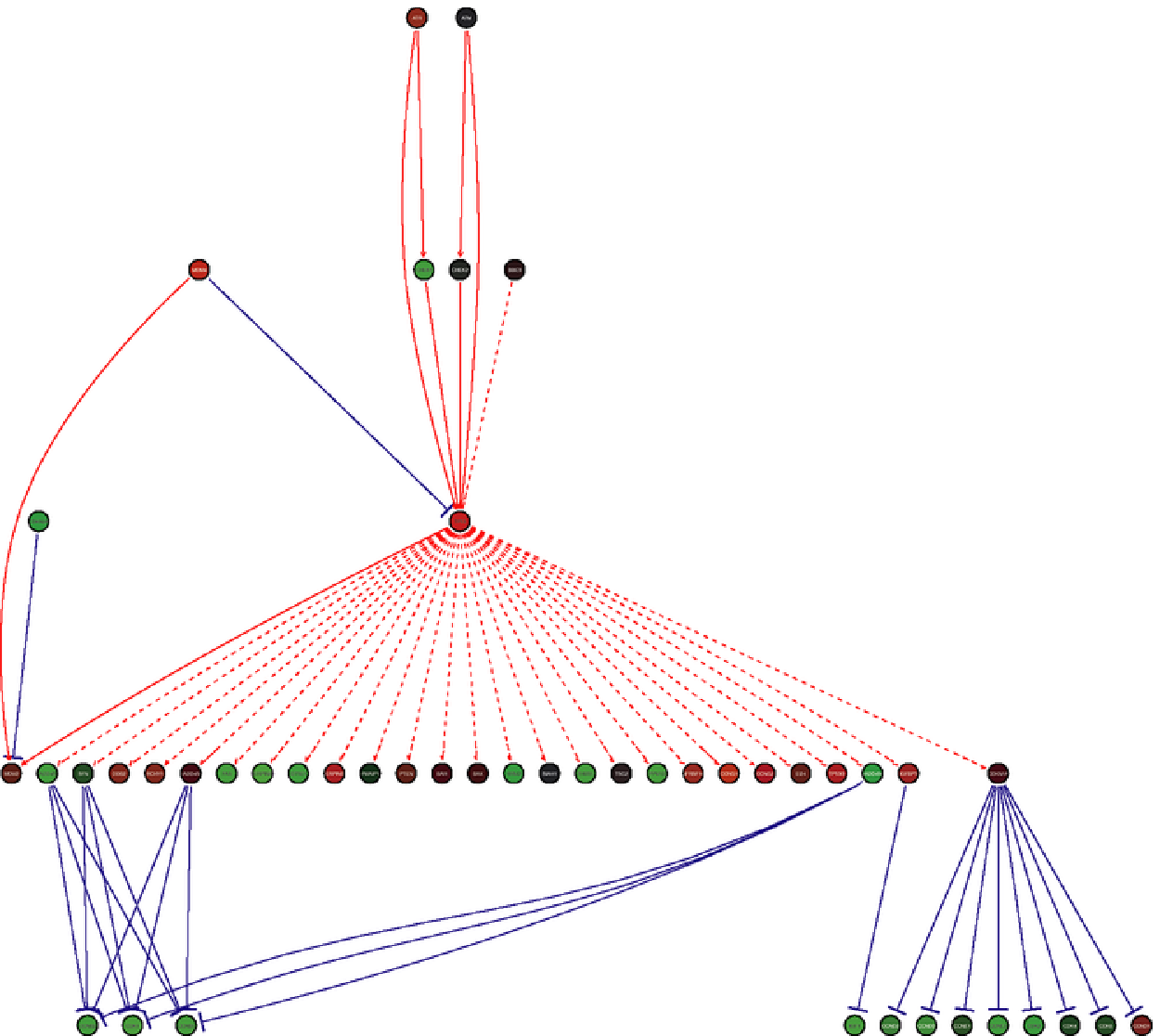}%
\vspace*{-3pt}
\caption{{Bladder cancer data set: KEGG p53 signaling pathway.}
Scaled difference in sample mean expression measures between T2$+$
and TaT1 tumors, for genes in one component of the KEGG p53
signaling pathway. Nodes are colored according to the value of the
difference in means, with green corresponding to high positive
values, red to high negative values, and black to $0$. Red arrows
denote activation, blue arrows inhibition.}
\label{fig:p53}\vspace*{-3pt}
\end{figure}

\begin{figure}

\includegraphics{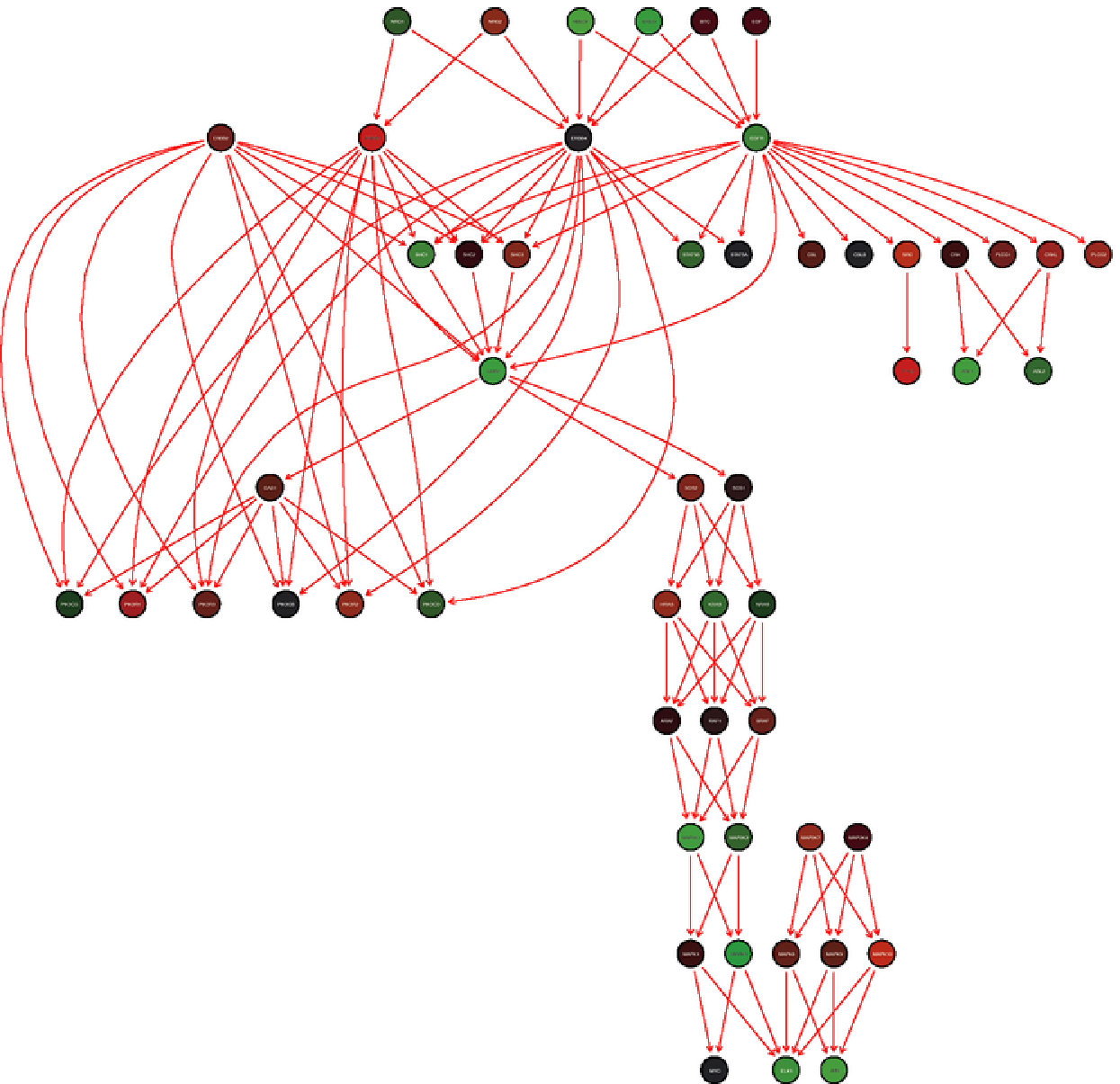}%
\vspace*{-4pt}
\caption{{Bladder cancer data set: KEGG ErbB signaling pathway.}
Scaled difference in sample mean expression measures between T2$+$
and TaT1 tumors, for genes in one component of the KEGG ErbB
signaling pathway. Nodes are colored according to the value of the
difference in means, with green corresponding to high positive
values, red to high negative values, and black to $0$. Red arrows
denote activation, blue arrows inhibition.}
\label{fig:erbb}\vspace*{-5pt}
\end{figure}

\textit{Bladder cancer data set}.
Testing the same KEGG networks on the bladder cancer data set, we
immediately notice that several gene sets which are well known to be
specific of one of the two bladder cancer progression pathways have
much lower $p$-values under our graph-based approach than using the
Hotelling $T^2$-statistic. This is the case, in particular, for the
\textit{p53 signaling
pathway} [\citet{Spruck1994Two},
\citet{Sanchez-Carbayo2006Defining}], which is
displayed in Figure~\ref{fig:p53} and for which the graph-based
procedure outputs a $p$-value of $8.5\times10^{-6}$ vs. $0.3$ for the
classical $T^2$-statistic. The TP53 gene itself is overexpressed in
invasive (T2$+$) tumors. \citet{Rhijn2004FGFR3} suggested that FGFR3 and
TP53 mutations characterize the two growth pathways and are mutually
exclusive. A more recent study [\citet{Hernandez2005FGFR3}] contradicts
the exclusion, but the observed underexpression of TP53 in the
invasive group could be coherent with its typical mutation in invasive
tumors. Genes coding for cyclins, such as CCNB1, CCNB2 and CDC2, are
overexpressed. Cyclins are positively involved in cell proliferation,
which is coherent with their overexpression in invasive tumors, as it
was already observed for other genes of the cyclin
family [\citet{Levidou2010D-type}]. IGF1 is also overexpressed in T2$+$
tumors, known to induce cell proliferation [\citet{Dunn1997Dietary}] and
was selected as a prognosis predictor for bladder
cancer in~\citet{Mitra2009Generation}.\looseness=-1

We also observe a much lower $p$-value using our procedure than using
the classical $T^2$-statistic ($2.3\times10^{-5}$ vs. $0.066$) for
the \textit{ErbB signaling pathway}, shown in Figure~\ref{fig:erbb} and
known to behave differently in the two bladder cancer growth
pathways [\citet{Mellon1996C-erbB-2}]. In particular, the network
involves the PIK3, RAS and MAPK genes, which are known to be
oncogenes specific to one of the growth
pathways [\citet{Eswarakumar2005Cellular}].

\begin{figure}

\includegraphics{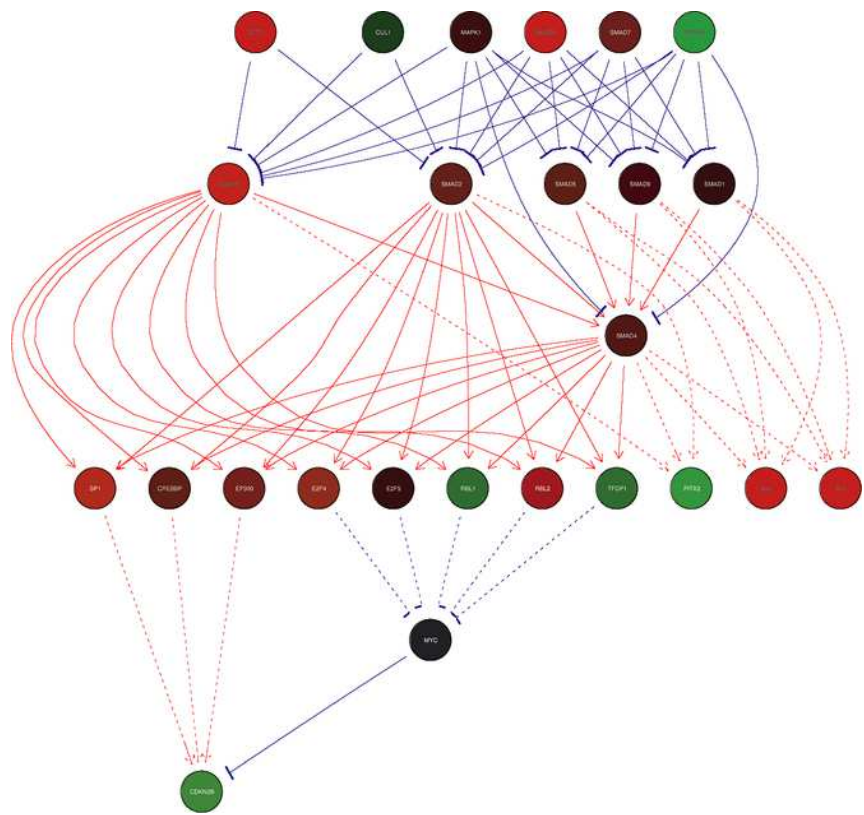}\vspace*{-3pt}

\caption{{Bladder cancer data set: KEGG TGF-$\beta$ signaling
pathway.}
Scaled difference in sample mean expression measures between T2$+$
and TaT1 tumors, for genes in one component of the KEGG
TGF-$\beta$ signaling pathway. Nodes are colored according to the
value of the difference in means, with green corresponding to high
positive values, red to high negative values, and black to
$0$. Red arrows denote activation, blue arrows inhibition.}\vspace*{-1pt}
\label{fig:tgfb}\vspace*{-3pt}
\end{figure}

Finally, changes in the \textit{TGF-$\beta$ signaling pathway} are also
known to be related\vadjust{\goodbreak} to the aggressiveness of bladder
cancers [\citet{Hung2008Molecular}]. The network is shown in
Figure~\ref{fig:tgfb} and here again our procedure results in a~much
lower $p$-value than the Hotelling test ($2.6\times10^{-6}$ vs.
$0.049$).

Unsurprisingly, these three networks have a relatively large size with
respect to the low sample size of this data set and several of their
genes show only very moderate differential expression when tested
individually.\vspace*{-3pt}

\subsection{NCI networks}
\label{sec:nci}

We also tested $75$ connected components coming from gene networks of
the NCI Pathway Integration%
Database.\footnote{\url{http://pid.nci.nih.gov}.} The NCI networks
considered are listed in the supplemental
article~\hyperref[app:pathwaytables]{Supplement B}
[\citet{Jacob2011More-SuppB}].
Unlike
KEGG pathways for which the Bioconductor R package \texttt{KEGGgraph}
had already been developed, NCI pathways were\vadjust{\goodbreak} not readily available as
R objects. We therefore developed
\texttt{NCIgraph} [\citet{Jacob2011NCIgraph}], a Bioconductor R package
which converts pathways available in BioPAX format to R objects. In
addition, instead of importing a gene network as is into R, we provide
an option to convert as well as possible the original network, whose
nodes can represent proteins, protein complexes or concepts like
transport or biochemical reactions, into one whose nodes correspond to
\textit{genes} and whose edges represent direct or indirect interactions
at the expression level. For instance, if protein A is known to
activate protein~B, which is a transcription factor for gene C, a~relevant network in terms of gene expression should be A and B
pointing to C, whereas the BioPAX network will most likely be
represented as~A pointing to B pointing to~C. As discussed in
Section~\ref{sec:synexpErr}, our method is robust to irrelevant edges
in the graph. Such a transformation is nonetheless important, since
the method essentially uses biological networks as a prior on the
covariance structure of gene expression. After this transformation,
however, most networks have much simpler topologies, typically with
all genes pointing to one or a few targets. As a result, Laplacian
eigenvalues often have high multiplicities, which makes the effect of
filtering less drastic.\footnote{If the eigenvalue $0$ has a very high
multiplicity, for example, then even the most extreme filtering
still retains a large number of dimensions.} In addition, the
networks we consider here have much smaller size than the KEGG
networks on average ($8.9$ vs. $36$ for means, $7.5$ vs. $23$ for
medians), which also explain the milder difference between results
before and after dimensionality reduction.\looseness=-1

For the breast cancer data, the $75$ connected components we consider
are those which have a nonempty intersection with the genes in this
microarray data set. As for the KEGG networks, we compare the classical
Hotelling $T^2$-test and the $T^2$-test in the new graph-based space
retaining only the first $20\%$ coefficients ($k=0.2 p$).

\begin{figure}

\includegraphics{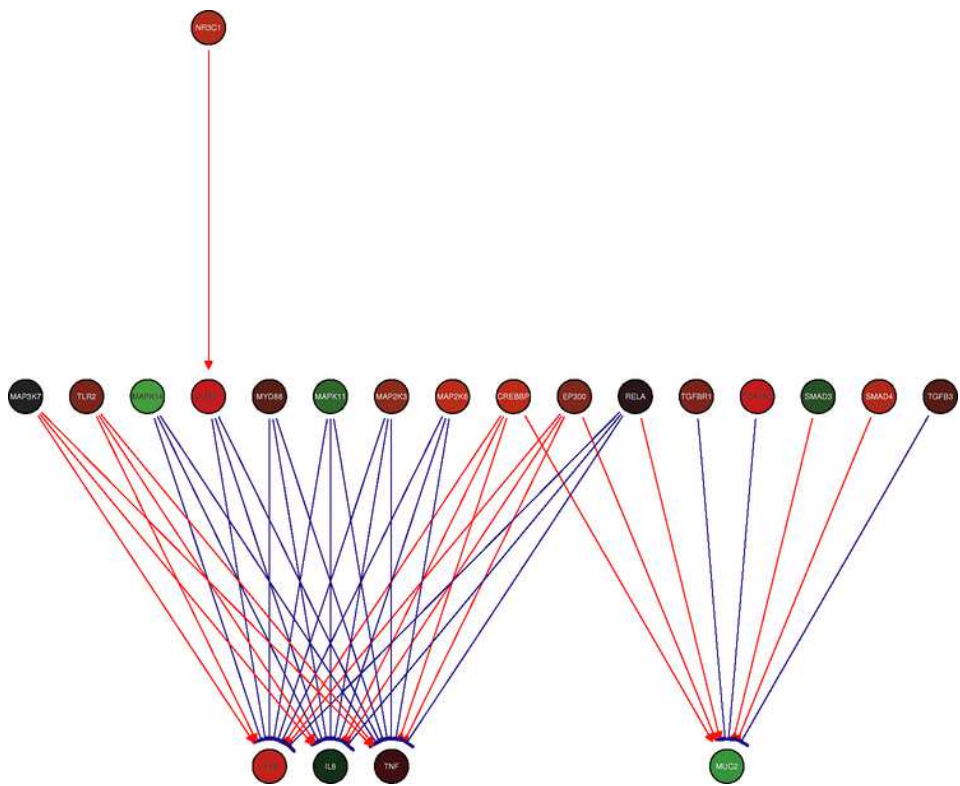}\vspace*{-3pt}
\caption{{Breast cancer data set: NCI Nfkb activation by nontypeable
hemophilus influenzae pathway.} Scaled difference in sample mean
expression measures between tamoxifen-resistant and sensitive
patients, for genes in one component of the NCI imported BioCarta
Nfkb activation by nontypeable hemophilus influenzae
pathway. Nodes are colored according to the value of the
difference in means, with green corresponding to high positive
values, red to high negative values, and black to $0$. Red arrows
denote activation, blue arrows inhibition.}\vspace*{-3pt}
\label{fig:nfkb}
\end{figure}

As an example, \textit{NFkB activation by Nontypeable Hemophilus
influenzae} shown in Figure~\ref{fig:nfkb} includes $21$ genes from
the breast cancer data set, but keeping the first $20\%$ of the
eigenvalues amounts to keeping $16$ dimensions because of
multiplicities. As a consequence, the $p$-value obtained after
filtering is only slightly lower than that before filtering. Here
again, the original context of study for this pathway has nothing to
do with breast cancer: the purpose was to uncover the inflammation and
mucin overproduction mechanism caused by a particular
bacteria. Nevertheless, this network contains several genes which are
either known actors of endocrine resistance or whose activity can be
directly linked to the resistance phenomenon. Moreover, as one may
expect, most of the observed gene-wise differential expression is
coherent with the annotated interactions. On the lower part of the
figure, IL1B is shown to be overexpressed in sensitive
patients. Consistent with this fact, its negative regulator p38
(MAPK11 and MAPK14) is downregulated in sensitive patients and its
positive regulator CREBBP is upregulated. Note that DUSP1\vadjust{\goodbreak} was
incorrectly annotated as a negative regulator in the automatic network
conversion process of \texttt{NCIgraph} but is actually a positive
regulator, as it is involved in the inactivation of p38. NR3C1 is
involved in the transcription of DUSP1 and is also upregulated in
sensitive patients. A few inconsistencies can be observed, like MAP2K3
and MAP2K6 which are negative regulators of IL1B, yet are
overexpressed in sensitive patients. Recall, however, that the criterion
we use for coherence is based on the difference between the expression
of each gene and the (interaction-sign corrected) average expression
of its regulators. The second main output of the pathway, MUC2, is
downregulated in sensitive patients, which makes sense both in terms
of the expression of its negative regulator TGFBR2, which is
upregulated, and the already observed
fact [\citet{Srinivasan2007Transcriptional}] that the estrogen receptor
upregulates MUC2 and that tamoxifen could block its
expression.\looseness=-1

The role of MUC2 in resistance to tamoxifen treatment of ductal
carcinoma does not seem to\vadjust{\goodbreak} be clearly established. Overexpression of
MUC2 is sometimes found to be mildly correlated with good
prognosis [\citet{Walsh1993Expression}, \citet{Rakha2005Expression}],
but this
may be caused by its correlation with ER$+$ status. Its overexpression
in resistant patients observed in this data set may well be noncausal,
but would deserve further investigation. As for TGFBR2, inactivating
mutations of the gene have been reported to be associated with
recurrence and tamoxifen resistance [\citet{Luecke2001Inhibiting}],
which is coherent with underexpression in resistant
patients. Regarding IL1B, the main output of the pathway, its
overexpression has been shown to be related to inhibition of cancer
growth through apoptosis [\citet{Roy2006Levels}]. DUSP1 is a known
negative regulator of cell proliferation and overexpression of p38 is
known to be related to tamoxifen
resistance [\citet{Gutierrez2005Molecular}]. Interestingly, NR3C1
activity has also been described by \citet{Wu2005Glucocorticoid} as
being related to breast cancer cell survival through its induction of
MAPK1 expression, which illustrates the interest of studying
differential expression patterns at a system level rather than at the
single-gene level.\looseness=-1

It is also important to note that at least two interpretations can be
given for the fact that sensitive patients have several gene
expression patterns corresponding to known factors of good
prognosis. Some of these patterns may be caused by the treatment, in
which case understanding how tamoxifen affects these genes in some
patients and not in others may be a proxy to understanding resistance
mechanisms. Some of the patterns though may also have been caused by
phenotypic traits of the sensitive patients, leading to better
prognosis but without any link to the treatment.

\begin{figure}

\includegraphics{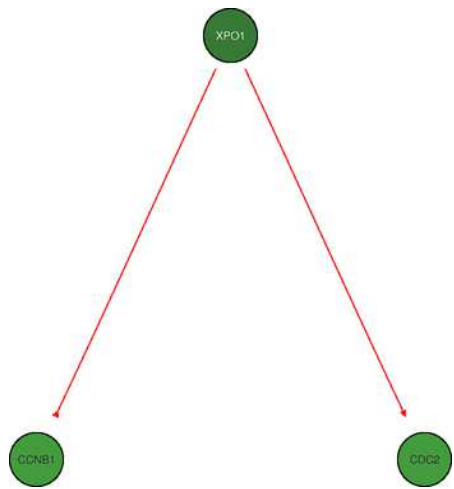}\vspace*{-5pt}
\caption{{Breast cancer data set: NCI sonic hedgehog receptor
ptc1 regulates cell cycle pathway.} Scaled difference in sample
mean expression measures between tamoxifen-resistant and sensitive
patients, for genes in one component of the NCI imported BioCarta
sonic hedgehog receptor ptc1 regulates cell cycle pathway. Nodes are
colored according to the value of the difference in means, with
green corresponding to high positive values, red to high negative
values, and black to $0$. Red arrows denote activation, blue
arrows inhibition.}
\label{fig:sonic}\vspace*{-6pt}
\end{figure}

Another small but relevant example is the \textit{sonic hedgehog
receptor ptc1 regulates cell cycle} pathway shown in Figure
\ref{fig:sonic}, which is entirely overexpressed in resistant patients,
yielding a 10-fold change between the $p$-value with and without
dimensionality reduction. The genes in this pathway are known to be
related to tamoxifen resistance: CCNB1 is related to proliferation and
is part of several existing tamoxifen-resistance signatures
[\citet{Paik2004Multigene}] and inhibition of CDC2 was already proposed
as an alternative treatment for endocrine resistant tumors
[\citet{Johnson2010Pre-clinical}].

\begin{figure}

\includegraphics{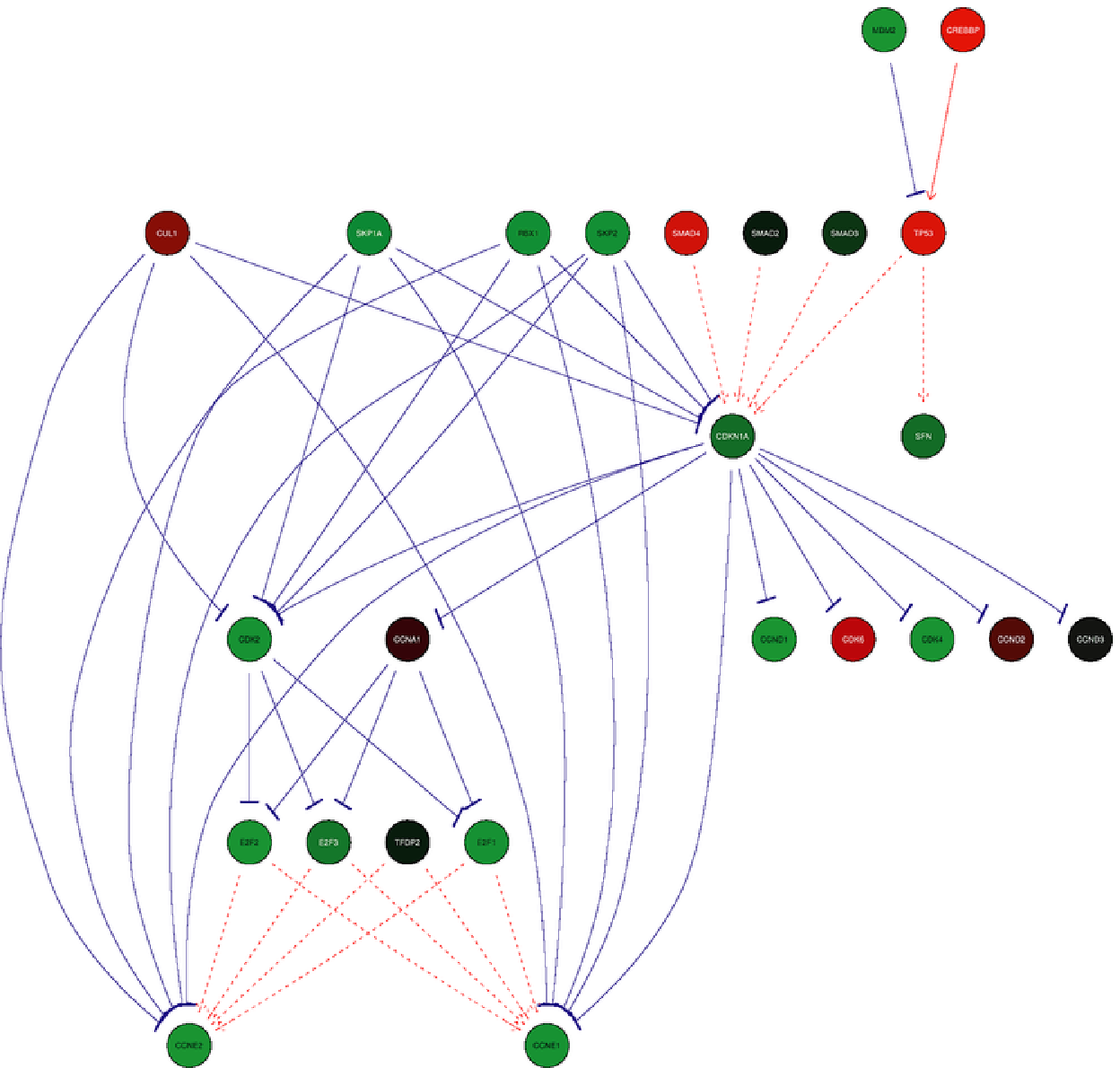}\vspace*{-5pt}
\caption{{Breast cancer data set: Subgraph discovery.} Difference
in sample mean expression measures between tamoxifen-resistant and
sensitive patients, for genes in the two overlapping subgraphs
detected at $\alpha=10^{-4}$. Nodes are colored according to the
value of the difference in means, with green corresponding to high
positive values, red to high negative values, and black to
$0$. Red arrows denote activation, blue arrows inhibition.}
\label{fig:firstGraphSign}\vspace*{-6pt}
\end{figure}

\subsection{Branch-and-bound subgraph discovery}

$\!\!\!$We ran our branch-and-bound nonhomogeneous subgraph discovery
procedure on the cell cycle pathway, whose largest connected
component, after restriction to edges of known sign (inhibition or
activation), has $86$ nodes and $442$ edges. Specifically, we sought
to detect differentially expressed subgraphs of size $q=5$, after
preselecting those for which the squared Euclidean norm of the
empirical shift exceeds $\theta=0.1$; for a test in the first $k=3$
components at level $\alpha=10^{-4}$, this corresponds to
$\lambda_{\min}<0.23$ and to an expected removal of $95\%$ of the
subgraphs under the approximation that the squared Euclidean norm of
the subgraphs follows a $\chi^2_5$-distribution.\vadjust{\goodbreak}

For $\alpha=10^{-4}$, out of $100$ runs on permuted data, only $9$
rejected the null hypothesis for at least one subgraph. More
precisely, $4$ of these $9$ runs detected $1$ subgraph and the others
detected $3$, $6$, $6$, $21$ and $26$ subgraphs. In contrast, $41$
overlapping subgraphs (Figure~\ref{fig:firstGraphSign}) were detected
on the original data, corresponding to a connected subnetwork of $25$
genes. Some of the genes belonging to these networks exhibit large
individual differential expression, namely, TP53 whose mutation has
been long known to be involved in tamoxifen resistance
[\citet{Andersson2005Worse}, \citet{Fernandez-Cuesta2010p53}].
Accordingly, its
negative regulator MDM2 is overexpressed and its positive regulator
CREBBP is underexpressed. E2F1, whose expression level was recently
shown to be involved in tamoxifen resistance
[\citet{Louie2010Estrogen}], is also part of the identified network, as
well as CCND1 [\citet{Barnes1997Cyclin}, \citet
{Musgrove2009Biological}]. Some
other genes in the network have quite low $t$-statistics and would not
have been detected individually. This is the case of CCNE1 and CDK2,
which were also described in \citet{Louie2010Estrogen} as part of the
same mechanism as E2F1. Similarly, CDKN1A was recently found to be
involved in anti-estrogen treatment resistance
[\citet{Musgrove2009Biological}] and in ovarian cancer, which is also a
hormone-dependent cancer [\citet{Cunningham2009Cell}]. Interestingly,
RBX1, a gene coding for a RING-domain E3 ligase known to be involved
in degradation of estrogen receptor $\alpha$ (ER$\alpha$)
[\citet{Ohtake2007Dioxin}], appears to be overexpressed in resistant
patients. This fact may suggest that some of the resistant\vadjust{\goodbreak} ER$+$
patients had fewer receptors and, as a result, their tumors were
relying less on estrogen for their growth, hence, the limited effect
of a~selective estrogen receptor modulator (SERM) like tamoxifen. The
networks also contain CDK4, whose inhibition was described in
\citet{Sutherland2009CDK} as acting synergistically with tamoxifen or
trastuzumab. More generally, a~large part of the network displayed in
Figure~2A of \citet{Musgrove2009Biological} is included in our network,
along with other known actors of tamoxifen resistance. Admittedly,
selecting an important regulator like TP53 is not a surprising result,
but our system-based approach to pathway discovery directly identifies
an important set of interacting genes and may therefore prove to be
more efficient than iterative individual identification of single
actors.\vadjust{\goodbreak}

\section{Software implementation}
\label{sec:soft}

The graph-structured test of Section \ref{sec:test} is implemented in
the R software package \texttt{DEGraph}, released through the
Bioconductor Project (release 2.7). Instructions for download and
installation are available at \url{http://www.bioconductor.org}. Note
that implementations of the branch-and-bound algorithms are not yet
included in this package, but are available upon request.

As mentioned in Section~\ref{sec:nci}, we also developed
\texttt{NCIgraph} [\citet{Jacob2011NCIgraph}], a~Bioconductor R package
which converts pathways available in BioPAX format to R objects with
various preprocessing options.\vspace*{-3pt}

\section{Discussion}
\label{sec:discussion}

$\!\!\!$We developed a graph-structured two-sample test of means, for problems
in which the distribution shift is assumed to be smooth on a~given
graph. We proved quantitative results on power gains for such
smooth-shift alternatives and devised branch-and-bound algorithms to
systematically apply our test to all the subgraphs of a large graph,
without enumerating and testing these subgraphs one-by-one. The first
algorithm is exact and reduces the number of explicitly tested
subgraphs. The second one is approximate, with no false positives and
a quantitative result on the type of false negatives (with respect to
the exact algorithm). The nonhomogeneous subgraph discovery method
involves performing a large number of tests, with highly-dependent
test statistics. However, as the actual number of tested hypotheses
is unknown, standard multiple testing procedures are not directly
applicable. Instead, we use a permutation procedure to estimate the
distribution of the number of false positive subgraphs. Such
resampling procedures (bootstrap or permutation) are feasible due to
the manageable run-time of the pruning algorithms of Section
\ref{sec:discovery}. Results on synthetic data illustrate the good
power properties of our graph-structured test under smooth-shift
alternatives, as well as the good performance of our
branch-and-bound-like algorithms for subgraph discovery. Very
promising results are also obtained on the gene expression data sets of
\citet{Loi2008Predicting} and \citet{Stransky2006Regional}.

Future work should investigate the use of other bases, such as
graph-wavelets [\citet{Hammond2009Wavelets}], which would allow the
detection of shifts with spatially-located nonsmoothness, for
example, to take into account errors in existing networks. As for the
cutoff selection, more systematic procedures should be considered,
{for example}, the two-step method proposed in~\citet{Das1974Power},
adaptive approaches as in~\citet{Fan1998Test} or heuristics based on
the eigengap as mentioned in Section~\ref{sec:geneexp}. The pruning
algorithm would naturally benefit from sharper bounds. Such bounds
could be obtained by controlling the condition number of all
covariance matrices, using, for example, regularized statistics which
still have known nonasymptotic distributions, such as those
of~\citet{Tai2008On}. Concerning multiple testing, procedures should
be devised to exploit\vadjust{\goodbreak} the dependence structure between the tested
subgraphs and to deal with the unknown number of tests. The proposed
approach could also be enriched to take into account different types
of data, {for example}, copy number for the detection of DE gene
pathways. More subtle notions of smoothness, {for example}, ``and''
(resp., ``or'') logical relations [\citet{Vaske2010Inference}], could
also be included to represent regulation mechanisms where the
simultaneous presence of two transcription factors (resp., the presence
of one or the other) is necessary to activate the transcription of
another gene. Other applications of two-sample tests with smooth-shift
on a graph include fMRI and eQTL association studies. For fMRI data,
the goal would be to detect whether the brain activity changes between
two conditions, using the prior information that parts of the brain
which are close up to brain convolutions or known connection patterns
should exhibit the same kind of change. One could also want to
identify specific areas of the brain whose activity changes between
two conditions. In eQTL studies, people are often interested in
finding genes whose expression is influenced by single-nucleotide
polymorphisms (SNPs), resulting in a large number of individual tests
which often need to be aggregated \textit{a posteriori} at the pathway
level. Our method could be used to identify pathways whose expression
is associated with particular SNPs.

Finally, it would be of interest to compare our testing approach with
structured sparse learning (which we briefly described in
Section~\ref{sec:intro}) for the purpose of identifying expression
signatures that are predictive of drug resistance. Methods should be
compared in terms of prediction accuracy and stability of the selected
genes across different data sets, a central and difficult problem in
the design of such
signatures [\citet{Ein-Dor2005Outcome},
\citet{He2010Stable},
\citet{Haury2010Increasing},
\citet{Haury2011influence}].\looseness=-1\vspace*{-3pt}

\section*{Acknowledgments}
$\!\!\!$The authors thank Anne Biton, Noureddine El Karoui, Za\"{i}d
Harchaoui, Miles Lopes and Terry Speed for very helpful discussions
and suggestions. They also acknowledge the UC Berkeley Center for
Computational Biology Genentech Innovation Fellowship, the Stand Up To
Cancer Program and The Cancer Genome Atlas Project for funding.\vspace*{-3pt}


\begin{supplement}
\sname{Supplement A}\label{app:technical}
\stitle{Technical results and proofs\\}
\slink[doi,text={10.1214/11-\break AOAS528SUPPA}]{10.1214/11-AOAS528SUPPA}
\slink[url]{http://lib.stat.cmu.edu/aoas/528/suppA.pdf}
\sdatatype{.pdf}
\sdescription{This section contains our technical results (Lemma and
Corollaries) on gain in power along with their proofs. It also
contains the upper bound used in the branch and bound algorithm
with its proof. Finally, it contains the lemma characterizing the
subgraphs that would be missed by the approximated subgraph
discovery algorithm presented in Section~\ref{sec:euclidean} along
with its proof.\vspace*{-3pt}}
\end{supplement}

\begin{supplement}
\sname{Supplement B}\label{app:pathwaytables}
\stitle{Pathways considered in the experiments\\}
\slink[doi]{10.1214/11-AOAS528SUPPB}
\slink[url]{http://lib.stat.cmu.edu/aoas/528/suppB.pdf}
\sdatatype{.pdf}
\sdescription{This section lists the names of the pathways
considered in the experiments.\vadjust{\goodbreak}}
\end{supplement}

\begin{supplement}
\sname{Supplement C}\label{app:glist}
\stitle{Gene lists}
\slink[doi]{10.1214/11-AOAS528SUPPC}
\slink[url]{http://lib.stat.cmu.edu/aoas/528/suppC.pdf}
\sdatatype{.pdf}
\sdescription{This section lists the genes belonging to each of the
pathways studied in detail in the experiments along with their
$t$-statistic and corresponding $p$-value.}
\end{supplement}

%

\printaddresses

\end{document}